\documentclass{iopjournal}

\usepackage{amssymb}
\usepackage{graphicx}
\usepackage{multirow}
\usepackage{dcolumn}
\usepackage{bm}
\usepackage{aas_macros}

\usepackage{natbib}

\usepackage{float}
\usepackage{xcolor}
\usepackage{subcaption}

\usepackage{array, booktabs, makecell}
\usepackage{siunitx, mhchem}
\usepackage[normalem]{ulem}

\newcommand{\OO}{\mathcal{O}}
\newcommand{\oh}{\frac{1}{2}}

\newcommand{\Eqref}[1]{Eq.\,(\ref{#1})}
\newcommand{\figref}[1]{Fig.\,\ref{#1}}
\newcommand{\tableref}[1]{Tab.\,\ref{#1}}
\newcommand{\secref}[1]{Sect.\,\ref{#1}}

\newcommand{\msol}{M_{\odot}}

\newcommand{\gccm}{\textrm{g\,cm}^{-3}}

\newcommand{\kapp}{\kappa'}

\def\tsc#1{\csdef{#1}{\textsc{\lowercase{#1}}\xspace}}
\tsc{WGM}
\tsc{QE}

\begin{document}

\articletype{Article type}

\title{Time integration for neutrino radiation transport using minimally implicit Runge-Kutta methods}

\author{Samuel Santos Pérez$^{1,*}$\orcid{https://orcid.org/0000-0002-8874-5546}, Martin Obergaulinger$^2$\orcid{0000-0001-5664-1382} and Isabel Cordero-Carrión$^{1}$\orcid{0000-0002-1985-1361}}

\affil{$^1$Departament de Matemàtiques, Universitat de València, carrer Dr. Moliner 50, 46100 Burjassot (València), Spain}

\affil{$^2$Departament d'Astronomia i Astrof\'isica, Universitat de Val\`encia, carrer Dr. Moliner 50, 46100, Burjassot (València), Spain}

\affil{$^*$Author to whom any correspondence should be addressed.}

\email{samuel.santos@uv.es}


\begin{abstract}
The evolution of many astrophysical systems is dominated by the interaction between matter and radiation such as photons or neutrinos. The dynamics can be described by the evolution equations of radiation hydrodynamics in which reactions between matter particles and radiation quanta couples the hydrodynamic equations to those of radiative transfer \citep[see][]{munier1986I,munier1986II}. The numerical treatment has to account for their potential stiffness (e.g., in optically thick environments). In this article, we will present a new method to numerically integrate these equations in a stable way by using minimally implicit Runge-Kutta methods. With these methods, the inversion of the implicit operator can be done analytically, so the computational cost is equivalent to that of an explicit method. In deriving the methods, we explicitly account for the behavior of the evolved variables in the stiff regime. We will show the results of applying these methods to the reactions between neutrinos and matter in some tests and also in realistic core-collapse supernovae simulations.
\end{abstract}

\section{Introduction} \label{sec:intro}

Radiation plays a crucial role in many fields of astrophysics. Besides representing the most important channel for observations of astronomical objects, its interaction with matter shapes their structure and drives their dynamics in quiescent as well as highly dynamic phases. Various types of reactions such as emission, absorption, and scattering exchange energy and momentum between radiation quanta and matter particles.  The corresponding cross sections and interaction rates can depend in very complicated ways on the thermodynamic properties of the gas and the radiation spectra. The resulting mean free paths of radiation can be very small compared to characteristic structural length scales (e.g., density scale heights) in optically thick regions with, for example, high gas densities, and at the same time exceeding the dimensions of the system in transparent regions such as atmospheric layers.  While the former limit is commonly reached for photons in the interior of ordinary stars, densities close to that of nuclear matter are required to turn matter optically thick to neutrino radiation. Thus, neutrinos are extremely important in core-collapse supernovae (CCSNe) and neutron stars, where this condition is met. Their importance is such that the explosion mechanism of CCSNe cannot be understood without a detailed account of the generation and transport of neutrinos.

During most of their lives, stars are in an equilibrium between their self-gravity and the thermal and radiation pressure that is ultimately provided by the nuclear reactions occurring in the stellar interior and converting hydrogen first to helium and then to heavier nuclei. Depending on the initial mass of a star, the sequence of reactions can reach the most tightly bound nuclei in the iron group. At that point, nuclear reactions cease to produce energy and the star loses stability against gravity, leading to the collapse of its core to a proto-neutron star (PNS). While the previous phases can be described as a sequence of hydrostatic equilibrium states, the final core collapse and the CCSN explosion that it can lead to are highly dynamic events and have to be modelled hydrodynamically. At all evolutionary stages, radiation and its interaction with matter are crucially important for the structure and evolution. At first, and generally in the outer layers, photons are the dominant form of radiation, but neutrinos end up taking over that role in the increasingly hot and dense cores. The interior of the PNS is optically thick even to neutrinos, which therefore diffuse out only slowly and transition to free streaming at the neutrinosphere near the PNS surface. Neutrinos are produced in the core by neutral and charged current reactions, with the latter being responsible for the deleptonisation of the PNS and the intense burst of electron neutrinos when the PNS forms. Some of the emitted neutrinos may be absorbed in the layers surrounding the PNS and heat the gas, which process is crucial for launching and powering the explosion that otherwise would fail due to the loss of energy in neutrinos

The intense coupling between radiation and matter implies that these events have to be modelled in the framework of (neutrino) radiation-hydrodynamics. The aforementioned high optical depth is a direct consequence of the very frequent interactions between neutrinos and matter. As the mean time between emission, absorption, or scattering reactions decreases far below the typical dynamical timescales of the core, the resulting stiffness adds another layer of complexity to the problem of solving the radiation-hydrodynamics equations, which is already challenging due to the multi-dimensional nature, the development  of turbulence, and the dependence of the reaction rates on neutrino energies. Besides these numerical challenges, the study of CCSNe is beset by physical uncertainties about, e.g., the nuclear equation of state or the way in which different neutrino flavours can transform into each other in the dense environment. Despite these limitations, the field of CCSN simulations has reached a certain level of maturity in which different groups using different simulation codes agree on basic features such as the neutrino burst, the evolution of the neutrino luminosities over the subsequent phases, or the role of neutrinos in the explosion mechanism briefly outlined above \citep[for recent reviews, see, e.g., ][]{Mueller2020LRR, Mezzacappa2020, Burrows2021Nature, Janka2025ARNPS}.

Much of the complexity of theoretically modelling the aforementioned systems comes from the equation underlying radiative transfer. At a basic level, the Boltzmann equation describes how the distribution function $f$ of radiation quanta evolves in a seven-dimensional phase space, consisting of time, position, and the momentum of the neutrino or photon. It takes into account transport terms involving divergence operators in position or momentum space as well as collision terms representing the interaction of one quantum with others or with matter that take the form of integrals over momentum space. As a rigorous treatment of the Boltzmann equation is feasible only in special cases, numerical methods used to model, e.g., CCSNe rely on approximations. In a very common approach, a momentum-space integration of the distribution function multiplied by the tensorial product of $n = 0, 1,...$ unit vectors yields a series of moments, now only functions of space and time. The first ones have a direct physical interpretation: $n = 0, 1,$ and $2$ correspond to the radiation energy density, momentum density and radiation pressure, respectively. The resulting infinite series of evolution equations takes the form of conservative form, in which the moment of order $n+1$ appears as a (spatial) flux of the moment of degree $n$ and source terms accounting for the reactions follow from moments of the collision integrals. We note that, when coupling radiative transfer to the dynamics of the gas to form the system of radiation hydrodynamics, the latter appears with an opposite sign in the fluid equations. Truncating the series at a finite degree $n$ and closing the system with a local algebraic relation for the higher moment(s) defines the family of the so-called $M_n$ methods. Other very different numerical approaches can be considered; for example, Monte Carlo methods can be used to include neutrino transport in the context of CCSNe and this strategy has been implemented in the general relativistic code SpEC \citep{Foucart2018, Foucart2021}.

$M_0$ or (flux-limited) diffusion and $M_1$ or algebraic Eddington tensor methods offer a good compromise between accuracy and numerical costs, and are thus widely used in relativistic astrophysics. $M_1$ methods are very good at modelling radiation in the optically thick and transparent regimes and also work well in the intermediate, semi-transparent regime. Nonetheless, a few difficulties remain. A particularly important one pertains to the time integration in the optically thick regime, in which the typical time scales of interactions between radiation and matter (the inverse of the reaction rates) can be many orders of magnitude smaller than the time scales associated to the radiation propagation or the dynamical time scales: the equations become stiff. A similar behaviour can be found in other conservation laws in many physical problems such as resistive relativistic magnetohydrodinamics \citep{cordero2023}, general relativistic force-free electrodynamics \citep{Mahlmann2021}, rarefied gases problems \citep{koellermeier2022rar} or shallow water equations \citep{koellermeier2022shallow}.

Designing methods for stiff equations requires specific considerations. Explicit time integration is only stable if the numerical time step is reduced to the characteristic time scales of the fastest evolving term, which in this case would be the radiation-matter interaction ones. Implicit methods, on the other hand, allow for a stable evolution even when using the --much larger-- time steps set by, e.g., radiation propagation or hydrodynamics. However, they can be very complicated to implement due to the inversion of the operators involved, in particular for parallel execution, and suffer from low computational efficiency. As a compromise, Implicit-Explicit (IMEX) Runge-Kutta methods \citep{pareschi2005implicit} combine an implicit integration of only the stiff terms with an explicit integration of the rest of the equations. This strategy has been used very recently by \citep{Izquierdo2022}. Semi-implicit numerical schemes \citep{Foucart2016} have also been used very recently in neutron star mergers (see, for example \citep{Radice2022}).

We present in this manuscript a new numerical scheme, that we are going to refer to as minimally implicit Runge-Kutta (MIRK) scheme, in order to solve the $M_1$ neutrino-hydrodynamics equations to first and second order in time that preserves stability properties of implicit methods and, at the same time, has a computing speed similar to that of an explicit method. The scheme proposed is based on allowing implicit evaluations of only evolved variables and, in practice, requires a slight modification from an explicit method, making its implementation in current numerical codes quite direct. We exploit the relaxation form of the source terms and their asymptotic behavior in the stiff regime in the derivation of the methods. We implement the new solver in the neutrino-hydrodynamics code of \citep{just2015} as an alternative scheme.

We subject the new scheme to a sequence of tests, starting with a simple system in which we evolve only the $M_1$ equations of radiative transport in a fixed, uniform background without coupling to hydrodynamics. The first of these tests follows the diffusion of radiation through layers with a preset constant opacity. The existence of an analytic solution allows us to test the stability and assess the convergence of the method in regimes of different stiffness. The second test is a simplified toy model for the emission of neutrinos in a PNS. Suppressing gas velocities, we set up a density and temperature profile and prescribe simple opacity laws. Neutrinos are produced in processes that represent thermal reactions and propagate from the optically thick centre of the PNS through a semi-transparent layer outward. While the gas density remains constant, we allow for a change of the temperature according to the radiative loss and gains of energy. The system develops a structure that approximates very roughly that of a supernova core. The systems is already too complex for an analytic solution to exist, but it is a useful test case for to explore the properties of our new method. The final test consists of a fully coupled neutrino-hydrodynamics simulation of a stellar core collapse in spherical symmetry. In terms of the microphysics employed, i.e., nuclear equation of state, neutrino transport and reactions, this setup is identical to typical production runs of supernova models. It differs from them only by the assumption of spherical symmetry. Physically, this is a very serious restriction as it suppresses the development of non-radial hydrodynamic instabilities and prevents an explosion. However, the dimensionality does not affect the - local - neutrino-matter interactions. We thus consider this set of simulations a stringent and meaningful test of how the MIRK scheme preforms in simulations of real astrophysical events.

This article is structured as follows. In \secref{sec:neu} we describe the main features of the equations governing the dynamics of the neutrino transport within the $M_1$ radiative transfer scheme. In \secref{sec:num} we explain in detail how the MIRK methods, up to second order, can be used to numerically integrate these equations, including the stability analysis in the stiff limit. In \secref{sec:results} we present some numerical tests, and we also show some numerical successful simulations of stellar core-collapse in spherical symmetry using our proposed MIRK scheme, first and second order. In the last, a more complex case, we compare the results with some other numerical approaches previously used. Finally, in \secref{sec:conclusions} we present our conclusions.

\section{Equations for neutrino transport in M1} \label{sec:neu}

The basic variables of $M_1$ radiative transfer, the energy density, $E (t, \vec r, \omega)$, and the momentum density, $\vec F (t, \vec r, \omega)$, of the radiation field, are functions of time $t$, position $\vec r$ and particle energy $\omega$ \citep{munier1986I, munier1986II}. Owing to their conservative character, the corresponding evolution equations take the form of balance laws including the spatial transport and the redistribution across particle energies by differential operators. Exchange of energy and momentum with matter enters the equations via source terms that typically depend only on the local state of radiation field and the matter, but not on their derivatives. Since we will deal with the latter terms, we write the $M_1$ system using the following short-hand notation (see also \citep{just2015}):
\begin{eqnarray}
    \partial_t E  &=& S_E + C^{(0)},\label{eq:m1e} \\
    \partial_t F^i &=& S_F^i+C^{(1),i},\label{eq:m1f}
\end{eqnarray}
where the terms with spatial or energy derivatives, $S_E$ and $S_F^i$, are split off from the interaction source terms, $C^{(0)}$ and $C^{(1),i}$. The form of the interaction source terms depends on the choice of interactions and possible approximations used to describe them. We specialise to the important case of thermal emission and absorption and isotropic scattering. Then, the rates of energy and momentum exchange are proportional to the absorption and transport opacities, $\kappa_{\mathrm{a}}$ and $\kappa_{\mathrm{tra}}$, respectively:
\begin{eqnarray}    \label{eq:CthermalE}
    C^{(0)} & = & c \, \kappa_a (E_{\text{eq}}-E),  \\
    \label{eq:CthermalF}
    C^{(1),i} & =& - c \, \kappa_{\text{tra}} F^i.
\end{eqnarray}
In the previous equations the constant $c$ denotes the speed of light in vacuum. Eq.~(\ref{eq:CthermalE}) describes how matter emits radiation thermally with an equilibrium distribution $E_{\text{eq}}$ (e.g., Maxwell-Boltzmann for photons, Fermi-Dirac for neutrinos), and how it absorbs the local radiation energy. Eq.~(\ref{eq:CthermalF}) accounts for the transfer of momentum to the gas by means of absorption and scattering reactions. We note that the same terms appear with the opposite sign (and integrated over particle energy) as sources in the hydrodynamic equations for the gas.

We do not delve deeper into the detailed, potentially very complicated, dependence of the opacities on $E$ and $\vec F$ as well as on the composition and thermodynamic state of the gas because our method is valid for general opacity laws. Our main focus lies on the stiff, optically thick limit, in which the opacities are very high, $\kappa_{\mathrm{a,tra}} \gg 1$, and the interaction terms dominate over $S_E$ and $S_F^i$ in Eqs.~(\ref{eq:m1e}) and (\ref{eq:m1f}). Under these conditions, numerical difficulties arise due to the need to simultaneously follow all the terms with characteristic time scales that can differ by many orders of magnitude.

The physically correct stiff limit consists of $E$ approaching the equilibrium energy density $E_{eq}$.  Furthermore, Eq.~(\ref{eq:CthermalF}) indicates that high opacities will reduce $\vec F$ to zero. However, the precise manner in which $\vec F$ vanishes matters a lot for getting the correct solution. In a non-uniform radiation field, $\vec F$ has to approach the diffusion limit satisfying  $$\displaystyle \vec F \to \vec F_{\mathrm{diff}} = \frac{1}{3 \, c \, \kappa_{\mathrm{tra}}} \vec \nabla E.$$ While some $M_1$ methods \citep{Jin_Levermore__1996__JCP__Hyperbol_stiff_relaxation, Pons_Ibanez_Miralles__2000__MNRAS__hyperbol_radtrans, Jin_et_al__2000__SIAM_JNA__RT_Diff_relax, Audit_et_al__2002__astro-ph__hyp_RHD_closure} deal with this requirement by explicitly enforcing the diffusion flux for high optical thickness, others \citep{just2015} found that an appropriate treatment of the flux terms in $S_{F}$ is sufficient to reproduce the correct limit. In practise, approaches such as the one of \citep{just2015} allow us to offload the issue of the correct diffusion limit to the solution of $S_{F}$. As long as our method for $C^{(1),i}$ ensures that $\vec F$ vanishes in the optically thick limit in the absence of $S_{F}$, the coupled solution of $S_{F}$ and $C^{(1),i}$ will behave correctly.

We focus now on the numerical schemes that can be used to solve Eqs. \eqref{eq:m1e} and \eqref{eq:m1f}.

The time-integration strategy for the transport equations, \eqref{eq:m1e} and \eqref{eq:m1f}, is usually chosen based on a trade-off between stability, accuracy, and numerical costs.  These goals are somewhat at odds with each other: the most stable schemes, implicit time integrators, and the most accurate ones, high-order methods, are also the most expensive ones; furthermore, high-order implicit methods tend to be particularly complex. The difficulties are exacerbated when applying the integrators to terms involving spatial as well as temporal derivatives. For this reason, an operator-splitting approach is common in which the transport terms, $S_{E,F}$, the interaction terms, $C^{0,1}$, and, in the case of coupled radiation hydrodynamics, the flux and source terms of the hydrodynamics equations not connected to neutrino interactions, are treated separately using suitable methods.  In the applications we are mostly interested in, CCSNe, we follow the evolution of the system on the hydrodynamical time scales, which leads us to select an explicit time integrator for the latter group of terms. Furthermore, the maximum hydrodynamic flow and sound speeds are similar to the characteristic velocities of the neutrino transport terms, which allows us to use an explicit time integrator for them with roughly the same stability constraint on the time step. On the other hand, their stiffness makes an implicit time integration scheme the only feasible option for the interaction terms.

The IMEX strategy described above is commonly employed in neutrino-hydrodynamics codes in high-energy astrophysics. Among the proposed methods, we follow the one implemented by \citep{just2015}, whose discretised schematics we briefly summarise in the following. We denote the conserved variables of hydrodynamics (the densities of mass, momentum, energy) and of the neutrino radiation ($E, \vec F$), collectively as $u$ and $w$, respectively, and use superscripts $^n$ and $^{n+1}$ to indicate the states at discrete time steps $t^{n}$ and $t^{n+1} = t^{n} + \Delta t$, respectively. Then, our prescription to update the variables to the next time step is given by
\begin{eqnarray}
    \label{Gl:RHD-imex-u}
    (u^{n+1} - u^{n} ) / \Delta t & = &     \mathcal{L}_{\mathrm{hydro}} (u^{n})
    + \bar{\mathcal{L}}_{\mathrm{int}} (u^{n}, w^{n+1}), \\
    \label{Gl:RHD-imex-w}
    (w^{n+1} - w^{n} ) / \Delta t & = &  \mathcal{L}_{\mathrm{tr}} (w^{n})
    + \mathcal{L}_{\mathrm{int}} (u^{n}, w^{n+1}),
\end{eqnarray}
where we the symbols $\mathcal{L}_{\mathrm{hydro}}$ and $\mathcal{L}_{\mathrm{tr}}$ stand for the discretised operators including the fluxes and sources of hydrodynamics and the fluxes of neutrino transport, respectively. Without entering into further details, we note that they are evaluated explicitly with data of the previous time step, $t^{n}$. The neutrino-matter interactions, represented by the operator $\mathcal{L}_{\mathrm{int}}$, i.e., the discretised version of Eqs.~\eqref{eq:CthermalE} and \eqref{eq:CthermalF}, depend on both $w$ and $u$. Their dependence on the hydrodynamic variables is a result of both the opacities and the equilibrium energy density, $E_{\mathrm{eq}}$, being functions of the thermodynamic state of the gas. We note that its counterpart in the hydrodynamic equations, $\bar{\mathcal{L}}_{\mathrm{int}}$, can be computed once $\mathcal{L}_{\mathrm{int}}$, and thus presents no further complication.

A fully implicit treatment of $\mathcal{L}_{\mathrm{int}}$ in \eqref{Gl:RHD-imex-w} would entail evaluating all the variables it depends on at the new time step, i.e., setting $\mathcal{L}_{\mathrm{int}} (u^{n+1}, w^{n+1})$. The intricate dependence of $\kappa_{\mathrm{a, tra}}$ and $E_{\text{eq}}$ on $u$ makes this task computationally costly, which burden the numerical solution. This step would require in general an iterative process that needs multiple times the recovery of the primitive (thermodynamic) variables, in particular the temperature, from $u$, and that implies the inversion of non-linear relations also multiple times. 

We propose a new scheme, the MIRK methods, that minimizes the computation cost of the process of recovery of variables. Our alternative approach differs in that we only allow the conserved neutrino variables to be evaluated implicitly, but treat the hydrodynamic variables and the variables derived from them, opacities and equilibrium energy density, explicitly. This simple change permits preserving the stability properties and simultaneously reducing the computational cost to that of an explicit method, as there is no need to apply the recovery multiple times. In the next section we explain the method in detail. The approach implemented by \citep{just2015} can be viewed as a particular case of the MIRK method. It was implemented without having in mind the mathematical framework presented in the next section. Here we go beyond their method, providing arguments based on the behavior of the evolved variables and stability criteria at the stiff limit. This mathematical framework also allows for a higher order extension in comparison with the approach presented in \citep{just2015}.

\section{Numerical methods} \label{sec:num}

This section presents the equations of a general MIRK method at first and second order. The general expressions contain undetermined coefficients that we will choose adequately in order to guarantee a correct behaviour in the stiff limit regime.

\subsection{First order method}

The equations of a first order MIRK method for Eqs. \eqref{eq:m1e} and \eqref{eq:m1f} take the form
\begin{eqnarray}
    E^{n+1}&=&E^n+\Delta t\left[S_E^n+a\,c\,\kappa^n_a\,(E^n_{\text{eq}}-E^n)\right. + \left.(1-a)\,c\,\kappa^n_a\,(E^n_{\text{eq}}-E^{n+1})\right], \\   (F^i)^{n+1}&=&(F^i)^n+\Delta t \left[(S^i_F)^n-b\,c\,\kappa^n_{\text{tra}}\,(F^i)^n\right.
    -\left.(1-b)\,c\,\kappa^n_{\text{tra}}\,(F^i)^{n+1}\right],
\end{eqnarray}
where $a,b$ are arbitrary real coefficients that we will select later according to stability criteria. From previous equations, the explicit expressions for $E^{n+1}$ and $(F^i)^{n+1}$ can be derived easily; they can be cast in matrix form as:
\begin{equation}
\begin{split}
    \begin{pmatrix}
    E \\ F^i
    \end{pmatrix}^{n+1}
    = \begin{pmatrix}
    E \\ F^i
    \end{pmatrix}^n +
    \begin{pmatrix}
    \frac{\Delta t}{1+\Delta t\,\kappa^n(1-a)} & 0 \\
    0 & \frac{\Delta t}{1+\Delta t\,\kapp^n(1-b)}\delta^{ij} 
    \end{pmatrix}   
    \begin{pmatrix}
    S_E + \kappa (E_{\text{eq}}-E) \\
    S^i_F - \kapp \, F^i
    \end{pmatrix}^n,
\label{eq:mat1}
\end{split}
\end{equation}
where $\kappa := c \, \kappa_a$ and $\kapp := c \, \kappa_{\text{tra}}$. The conditions $a,b<1$ must be satisfied to force non-zero (and positive) denominators always. Notice that the equations in this form resemble a pure explicit method with effective time steps $$\displaystyle \Delta t_E=\frac{\Delta t}{ 1+\Delta t\,\kappa^n(1-a)},\;\;\displaystyle \Delta t _F=\frac{\Delta t}{ 1+\Delta t\,\kapp^n(1-b)}$$ for the $E$ and $F^i$ evolution equations, respectively. The previous matrix expression has been easily and analytically derived thanks to the fully explicit evaluation of the non conserved variables (e.g., all the variables different from $E$ and $F^i$). Due to this reason, one would expect to have a computational cost similar to that of applying a fully explicit method. We now analyze the behaviour in the stiff limit regime.

Mathematically speaking, the stiff limit refers to $\kappa_a, \kappa_{\text{tra}} \to \infty$. In that limit, Eq.~\eqref{eq:mat1} reads
\begin{equation}
    \begin{pmatrix}
    E \\ F^i
    \end{pmatrix}^{n+1}
    = \begin{pmatrix}
    \frac{-a}{1-a} & 0 \\
    0 & \frac{-b}{1-b}\delta^{ij} 
    \end{pmatrix}
    \begin{pmatrix}
    E \\ F^i
    \end{pmatrix}^n
    + \begin{pmatrix}
    \frac{ E_{\text{eq}}^n}{1-a} \\
    0
    \end{pmatrix}.
\end{equation}
Conditions
\begin{equation}
a<1/2,\quad b<1/2,
\label{eq:cond1}
\end{equation}
must be fulfilled for the spectral radius of the updated matrix to be strictly bounded by 1, a necessary condition to have a stable numerical method at the stiff limit. This is a more restrictive condition in comparison with previous conditions $a,b<1$ (needed to avoid zero values in the denominators).

In order to guarantee a correct behaviour of the numerical solution at the stiff limit, and assuming well-behaved and smooth data for the time step $n$ (this is $E^n=E_{\text{eq}}^n+\OO(\Delta t)$ and 
$(F^i)^{n}= 0 + \OO(\Delta t)$), we see that, for all $a, b$, for the time step $(n+1)$, it must be satisfied:
\begin{gather*}
E^{n+1}=E_{\text{eq}}^{n+1}+\OO(\Delta t),\\
(F^i)^{n+1}= 0 + \OO(\Delta t).
\end{gather*}
So, independently on the values of the coefficients $a, b$ we get well-behaved and smooth data in the next time step at first order. This gives us, in principle, full freedom for choosing $a$ and $b$. However, the behaviour of the evolved variables are far from been smooth in supernovae simulations and other astrophysical scenarios. Therefore, we should guarantee their correct behaviour at the stiff limit even when we are dealing with non-smooth data, and regardless the possible presence of numerical errors in the previous time steps. The choice $b=0$ guarantees the correct behaviour for $F$ at the stiff limit, i.e., $(F^i)^{n+1}=0$. It remains to choose a value for $a$. By analogy with $b$, and taking into account the particular case $E_{\text{eq}}=0$, we will simply consider $a=0$. This means that the behaviour of $E$ at the stiff limit is not controlled by previous values of this quantity, but only by evaluations of $E_{\text{eq}}=E_{\text{eq}}(u)$, which only depends on the hydrodynamic variables $u$. With this choice, it is satisfied that $E^{n+1} = E_{\text{eq}}^n = E_{\text{eq}}^{n+1}+\OO(\Delta t)$. Finally, in the case $a=b=0$ the method reads:
\begin{eqnarray}
    \label{eq:opthlim1-e}
    E^{n+1}&=&E^n+\frac{\Delta t}{1+\Delta t\,\kappa^n}\left[S_E^n + \kappa^n (E^n_{\text{eq}}-E^{n})\right], \;\;\;\;\;\;\\
    \label{eq:opthlim1-f}
    (F^i)^{n+1}&=&(F^i)^n+\frac{\Delta t}{1+\Delta t\,\kapp^n} \left[(S_F^i)^n - \kappa'^n (F^i)^n\right].\;\;\;\;
\end{eqnarray}

\subsection{Second order method}

Hereafter we follow the same strategy as in the first order case. Two stages are needed for the second order method. We denote the intermediate step by a $^{(1)}$ superscript and the final step by a $^{n+1}$ superscript. In general, we have four coefficients, $a, a', b, b'$, to be determined based on stability arguments. The first stage reads
\begin{eqnarray}
    E^{(1)} &=& E^n + \Delta t \left[ S_E^n + a \, \kappa^n (E^n_{\text{eq}}-E^n) \right.+ \left.(1-a) \, \kappa^n (E^n_{\text{eq}}-E^{(1)})\right],\label{eq:me12}\\
    (F^i)^{(1)} &=& (F^i)^n + \Delta t \left[ (S^i_F)^n - b \, \kappa'^n (F^i)^n \right. - \left. (1-b) \, \kappa'^n (F^i)^{(1)}\right],\label{eq:mf12}
\end{eqnarray}
and the second stage can be written as
\begin{eqnarray}
    E^{n+1}&=&\oh[E^{(1)}+E^n]+\Delta t\Big[\oh S_E^{(1)}+a'\,\kappa^{(1)}(E^{(1)}_{\text{eq}}-E^{(1)})\nonumber\\&&+\frac{1-a}{2}\kappa^{(1)}(E^{(1)}_{\text{eq}}-E^{n})+  \left(\frac{a}{2}-a'\right)\kappa^{(1)}(E^{(1)}_{\text{eq}}-E^{n+1})\Big], \\
   (F^i)^{n+1}&=&\oh[(F^i)^{(1)}+(F^i)^n]+\Delta t\,\Big[\oh(S^i_F)^{(1)}-b'\kappa'^{(1)}(F^i)^{(1)} \nonumber\\
   &&-\frac{1-b}{2}\kappa'^{(1)}(F^i)^n-\left(\frac{b}{2}-b'\right)\kappa'^{(1)}(F^i)^{n+1}\Big].
\end{eqnarray}
Isolating $E^{(1)}$ and $(F^i)^{(1)}$, we get similar expressions to those of the first order, just substituting the superscript $^n$ by $^{(1)}$:
\begin{equation}
    E^{(1)} = E^n
    +\frac{\Delta t}{1+\Delta t\,\kappa^n(1-a)}\left[S_E^n + \kappa^n (E^n_{\text{eq}}-E^{n})\right],\label{eq:me12i}
\end{equation}
\begin{equation}
    (F^i)^{(1)}=(F^i)^n
    +\frac{\Delta t}{1+\Delta t\,\kapp^n(1-b)} \left[(S_F^i)^n - \kappa'^n (F^i)^n\right].\label{eq:mf12i}
\end{equation}
Then, $E^{n+1}$ and $F^{i,n+1}$ can be expressed explicitly in terms of previous evaluations of these quantities as:
\begin{equation}
    \begin{split}
    E^{n+1}=& \left[1+\Delta t \,\kappa^{(1)}\left(\frac{a}{2}-a'\right)\right]^{-1} \bigg\{ 
    \left[\oh-\Delta t \,\kappa^{(1)}\left(\frac{1-a}{2}\right)\right] E^n + \left[\oh-\Delta t \,\kappa^{(1)}a'\right] E^{(1)} \\
    &+ \frac{\Delta t \, S_E^{(1)}}{2} +\frac{\Delta t\,\kappa^{(1)}E_{\text{eq}}^{(1)}}{2}\bigg\} \\
    =& \left[1+\Delta t \,\kappa^{(1)}\left(\frac{a}{2}-a'\right)\right]^{-1} \bigg\{ \left[\oh-\Delta t \,\kappa^{(1)}\left(\frac{1-a}{2}\right)\right] E^n + \left[\oh+\Delta t \,\kappa^{(1)}\left(\oh-a'\right)\right] E^{(1)} \\
    &+\frac{\Delta t}{2} \left[ S_E^{(1)} + \kappa^{(1)}(E_{\text{eq}}^{(1)}-E^{(1)})\right]\bigg\},
    \end{split}
\label{eq:me2i}
\end{equation}
\begin{equation}
    \begin{split}
    (F^i)^{n+1}=&\left[1+\Delta t\,\kapp^{(1)}\left(\frac{b}{2}-b'\right)\right]^{-1}\bigg\{ \left[\oh-\Delta t\,\kapp^{(1)}\left(\frac{1-b}{2}\right)\right] (F^i)^n \\
    &+ \left[\oh-\Delta t\,\kapp^{(1)}b'\right] (F^i)^{(1)} + \frac{\Delta t \, (S_F^i)^{(1)}}{2}\bigg\} \\
    =&\left[1+\Delta t\,\kapp^{(1)}\left(\frac{b}{2}-b'\right)\right]^{-1}\bigg\{ \left[\oh-\Delta t\,\kapp^{(1)}\left(\frac{1-b}{2}\right)\right] (F^i)^n + \left[\oh+\Delta t\,\kapp^{(1)}\left(\oh-b'\right)\right] (F^i)^{(1)} \\
    & +\frac{\Delta t}{2}\left[(S_F^i)^{(1)}-\kapp^{(1)}(F^i)^{(1)}\right]\bigg\}.
\end{split}
\label{eq:mf2i}
\end{equation}

The following conditions are necessary for ensuring non-zero (positive) denominators always:
\begin{equation}\frac{a}{2}-a'>0,\quad 1-a>0;\quad
\frac{b}{2}-b'>0,\quad 1-b>0.
\label{eq:cond2}
\end{equation}
We finally determine the coefficients of the method taking into account previous conditions and the behaviour of the numerical solution at the stiff limit.

The stiff limit refers to $\kappa^n,\kappa^{(1)},\kapp^{n},\kapp^{(1)}\to \infty$. In that limit, Eqs.~\eqref{eq:me2i} and \eqref{eq:mf2i} read
\begin{equation}
E^{n+1}=
\lambda_E E^n - \frac{a'}{\left(\frac{a}{2}-a'\right)(1-a)}E_{\text{eq}}^n + \frac{1/2}{\frac{a}{2}-a'}E_{\text{eq}}^{(1)},\label{eq:me2s}
\end{equation}
\begin{equation}
(F^i)^{n+1}=\lambda_F (F^i)^n,\label{eq:mf2s}
\end{equation}
where 
\begin{eqnarray}
    \lambda_E=\frac{a'a-\left(\frac{1-a}{2}\right)(1-a)}{\left(\frac{a}{2}-a'\right)(1-a)}, \\
    \lambda_F=\frac{b'b-\left(\frac{1-b}{2}\right)(1-b)}{\left(\frac{b}{2}-b'\right)(1-b)}.
\end{eqnarray}
$|\lambda_E|\leq 1$ and $|\lambda_F|\leq 1$ must be satisfied as a necessary condition to have stability of the numerical method at the stiff limit.

If we assume well-behaved and smooth data at second order in time at the stiff limit at $t=t^n$,
\begin{equation}
    E^n = E_{\text{eq}}^n+\OO(\Delta t^2), \quad
    (F^i)^n = 0 + \OO(\Delta t^2), \label{eq:efgd2}
\end{equation}
from Eqs.~\eqref{eq:me2s} and \eqref{eq:mf2s} we get
\begin{equation}
    E^{n+1} = E_{\text{eq}}^{n+1}
    + \frac{a'+\frac{1-a}{2}}{\frac{a}{2}-a'}\left(\nabla_X E_{\text{eq}}\right)^n\cdot \tilde{S}_X^n\Delta t +\OO(\Delta t^2), \label{eq:megdnp1}
\end{equation}
\begin{equation}
    (F^i)^{n+1} = 0 + \OO(\Delta t^2), \label{eq:mfgdnp1}
\end{equation}
where we have used a Taylor expansion over $E^{(1)}_{\text{eq}}$, $X$ represents the vector of all the evolved variables (i.e. $X=(u,w)$), $\nabla_X = (\partial_{X_1}, \partial_{X_2}, ... )$ and $\tilde{S}_X$ is the source term in the evolution equation of the form $\partial_t X=\tilde{S}_X$. So in order to satisfy Eqs.~\eqref{eq:efgd2} in the next time step, $t^{n+1}$, we need to impose $a'=\frac{a-1}{2}$. We could choose $b'$ in resemblance with $a'$, having then
\begin{equation}
a'=\frac{a-1}{2}, \quad b'=\frac{b-1}{2}. \label{eq:choiceb1}
\end{equation}
This choice for $a'$ preserves the second order behaviour of our numerical solution in the next time step and at the stiff limit when smooth data are involved, $a/2-a'>0$ is trivially satisfied and $\lambda_E=-1$; note that having $\lambda_E=-1$ means considering a value for $\lambda_E$ at the boundary of the linear stability region, which seems to be not very convenient. In addition, as commented for the first order method, we must take into account that the evolved variables have a non-smooth behaviour in supernovae simulations and other astrophysical scenarios. 

Our proposal is then to consider $b\neq 0$ and $b'=(b-1)^2/(2b)$ to guarantee $\lambda_F = 0$ (as we did choosing $b=0$ for the first order method), preserving the correct behaviour of our numerical solution regardless the possible presence of numerical errors or non-smooth data. By analogy with $b'$, and taking into account the particular case $E_{\text{eq}}=0$, we will consider $a\neq 0$ and $a'=(a-1)^2/(2a)$, or $\lambda_E=0$ equivalently. For these choices, the conditions \eqref{eq:cond2} result in 
\begin{equation}
    a'=\frac{(1-a)^2}{2a},\;\; a\in(-\infty,0)\cup (1/2,1),\label{eq:choiceb2a}
\end{equation}
\begin{equation}
    b'=\frac{(1-b)^2}{2b},\;\; b\in(-\infty,0)\cup (1/2,1),
    \label{eq:choiceb2}
\end{equation}
and Eq.~\eqref{eq:megdnp1} stays
\begin{equation}
    E^{n+1} = E_{\text{eq}}^{n+1}
    + \frac{1-a}{2a-1}\left(\nabla_X E_{\text{eq}}\right)^n\cdot \tilde{S}_X^n\Delta t +\OO(\Delta t^2).
\end{equation}
Second order for smooth data at stiff limit would be guaranteed if $a=1$, but this is incompatible with conditions \eqref{eq:choiceb2a}. We only get first order in time at the stiff limit for $E$, a price to pay when non-smooth data is considered. Within these constraints, we still have some freedom for the election of the coefficients $a,b$.

We can write our numerical method in such a way it resembles a pure explicit scheme of the form
\begin{gather}
    X^{(1)}=X^n+\Delta t \tilde{S}_X^n,
\label{eq:exp2is}\\
    X^{n+1}=\frac{X^n}{2}+\frac{X^{(1)}}{2}+\Delta t\frac{\tilde{S}_X^{(1)}}{2}.
\label{eq:exp2np1}
\end{gather}

For the choices \eqref{eq:choiceb1}, Eqs.~\eqref{eq:me12i} and \eqref{eq:mf12i} keep the same form, while Eqs.~\eqref{eq:me2i} and \eqref{eq:mf2i} can be written as:
\begin{equation}
\begin{split}
       E^{n+1}=&\frac{1-\Delta t\,\kappa^{(1)}(1-a)}{2+\Delta t\,\kappa^{(1)}}E^n +\frac{1+\Delta t\,\kappa^{(1)}(2-a)}{2+\Delta t\,\kappa^{(1)}}E^{(1)}
    +\frac{\Delta t}{2+\Delta t\,\kappa^{(1)}}\left[S^{(1)}_E+\kappa^{(1)}(E_{\text{eq}}^{(1)}-E^{(1)})\right],
\end{split}
\end{equation}
\begin{equation}
\begin{split}
    (F^i)^{n+1}=&\frac{1-\Delta t\,\kapp^{(1)}(1-b)}{2+\Delta t\,\kapp^{(1)}}(F^i)^n 
    +\frac{1+\Delta t\,\kapp^{(1)}(2-b)}{2+\Delta t\,\kapp^{(1)}}(F^i)^{(1)} 
    +\frac{\Delta t}{2+\Delta t\,\kapp^{(1)}}\left[(S_F^i)^{(1)}-\kappa'^{(1)}(F^i)^{(1)}\right], 
\end{split}
\end{equation}
where $a,b$ have not been chosen yet. For $\kappa,\kappa'\to 0$, we recover the structure of the previous second order pure explicit method, Eqs.~\eqref{eq:exp2is} and \eqref{eq:exp2np1}.

Finally, with conditions \eqref{eq:choiceb2a} and \eqref{eq:choiceb2}, Eqs.~\eqref{eq:me12i} and \eqref{eq:mf12i} keep the same form, while Eqs.~\eqref{eq:me2i} and \eqref{eq:mf2i} can be written as:
 \begin{equation}
     \begin{split}
         E^{n+1}=&\frac{1-\Delta t\,\kappa^{(1)}(1-a)}{2+\Delta t\,\kappa^{(1)}\left(2-\frac{1}{a}\right)}E^n + \frac{1+\Delta t\,\kappa^{(1)}\left(3-\frac{a^2+1}{a}\right)}{2+\Delta t\,\kappa^{(1)}\left(2-\frac{1}{a}\right)}E^{(1)}\\ 
         +& \frac{\Delta t}{2+\Delta t\,\kappa^{(1)}\left(2-\frac{1}{a}\right)}\left[S^{(1)}_E+\kappa^{(1)}(E_{\text{eq}}^{(1)}-E^{(1)})\right],
     \end{split}
 \end{equation}
 \begin{equation}
 \begin{split}
        (F^i)^{n+1}=&\frac{1-\Delta t\,\kapp^{(1)}(1-b)}{2+\Delta t\,\kapp^{(1)}\left(2-\frac{1}{b}\right)}(F^i)^n 
    + \frac{1+\Delta t\,\kapp^{(1)}\left(3-\frac{b^2+1}{b}\right)}{2+\Delta t\,\kapp^{(1)}\left(2-\frac{1}{b}\right)}(F^i)^{(1)} \\
    +& \frac{\Delta t}{2+\Delta t\,\kapp^{(1)}\left(2-\frac{1}{b}\right)}\left[(S_F^i)^{(1)}-\kappa'^{(1)}(F^i)^{(1)}\right], 
 \end{split}
\end{equation}
with $a,b$ still to be chosen. For $\kappa,\kappa'\to 0$, we also recover the structure of the previous second order pure explicit method.

\section{Numerical simulations and results} \label{sec:results}
\subsection{Diffusion limit test}
\label{sec:Pons1}
As a first test for the MIRK methods we deal with a simplification of the $M_1$ equations by imposing spherical symmetry and the diffusion limit (see \citep{Pons_Ibanez_Miralles__2000__MNRAS__hyperbol_radtrans} for more details). Doing so we get the following expressions associated to the $M_1$ equations:
\begin{equation}    S_E=\partial_rF+\frac{2F}{r}, \quad S_F=\frac{1}{3}\partial_rE, \quad \kappa_a=0, \quad \kappa_{tra}=\kappa,
\end{equation}
with $\kappa$ being a constant value (do not confuse with the notation used in the previous section). The $M_1$ system has therefore the following exact analytical solution:
\begin{equation}
    E(t,r)=\left(\frac{\kappa}{t}\right)^{3/2}\exp{\left(-\frac{3\kappa r^2}{4t}\right)}, \quad
    F(t,r)=\frac{r}{2t}E(t,r).
\end{equation}
The parameter values we are using in what follows for this model are taken from the ones used in the first test of \citep{Pons_Ibanez_Miralles__2000__MNRAS__hyperbol_radtrans}. The spatial domain will always be $r\in[0,1]$. A zero radial derivative and a zero value for the inner boundary conditions are applied to $E$ and $F$, respectively. At the outer boundary we consider a zero derivative for both variables. All spatial grids in this subsection begin at $r_0=10^{-5}$, to avoid numerical errors coming from the treatment of the inner boundary to be significant with lower resolutions (and faster simulations).

First, we set $\kappa=10^2$ and use a piece-wise linear reconstruction method for $S_{E,F}$ with a slope-limiter that employs the minmod function (see \citep{leveque1992numerical} for more details).

The time step $\Delta t$ is computed using a CFL factor as the minimum of the local admissible time steps,
\begin{equation}
    \Delta t = \mathrm{CFL} \min_{\mathrm{grid~cells}} \left\{ \frac{\Delta r}{|c_{\mathrm{max}}|} \right\},
\end{equation}
where $c_{\mathrm{max}}$ is the maximum characteristic speed in  each cell. In our, highly diffusive, case without fluid velocities, it is given by $|c_{\mathrm{max}}| = 1 / \sqrt{3}$, and thus independent of position, time, and resolution. Hence, the time step is given by the (uniform) spatial resolution and the CFL factor as $\Delta t = \sqrt{3} \, \Delta r \, \mathrm{CFL}$.

\begin{figure}[htbp!]
  \centering     \includegraphics[width=0.49\linewidth]{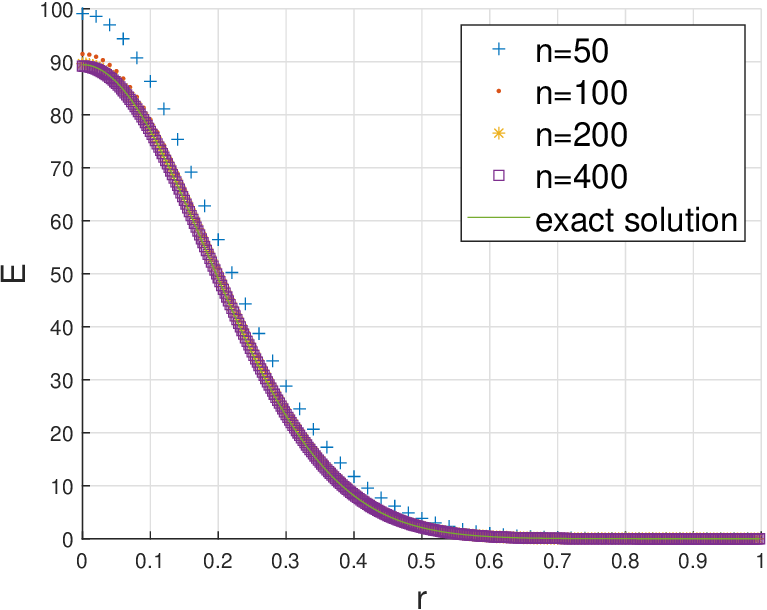}
      \includegraphics[width=0.49\linewidth]{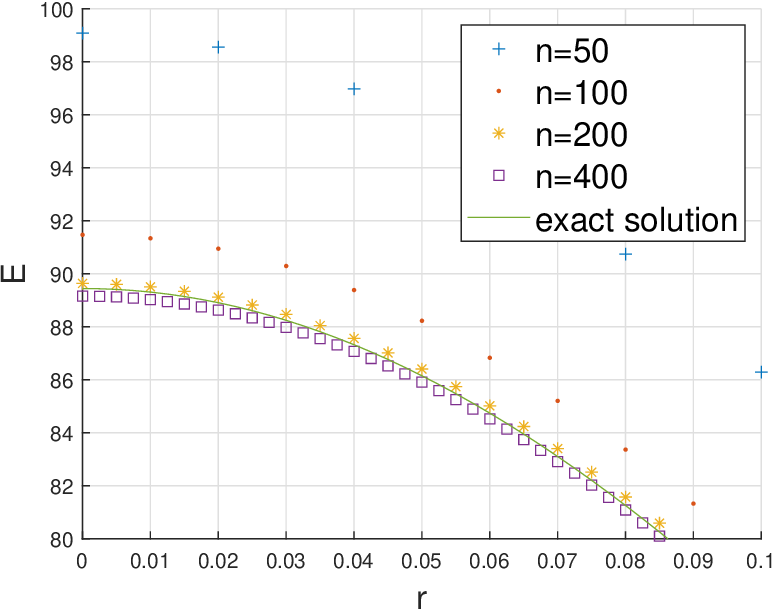}
  \caption{Results for $E$ using the MIRK1 method with $\kappa=10^2$ and CFL=$1/2^5$. Several spatial resolutions at $t=5$ are displayed. The whole spatial domain is shown at the left panel and a zoom near the origin is displayed at the right panel.}
  \label{fig:DLEk100mirk1}
\end{figure}

\begin{figure}[htbp!]
  \centering
      \includegraphics[width=0.49\linewidth]{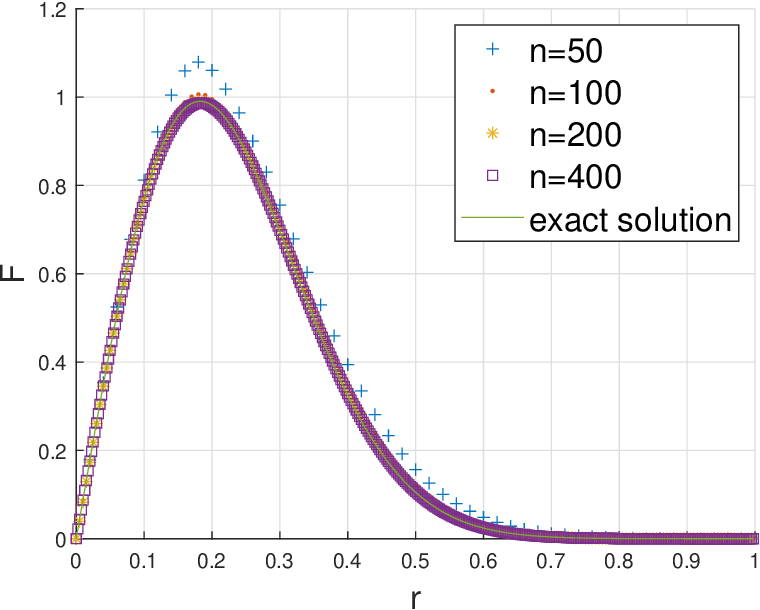}
      \includegraphics[width=0.49\linewidth]{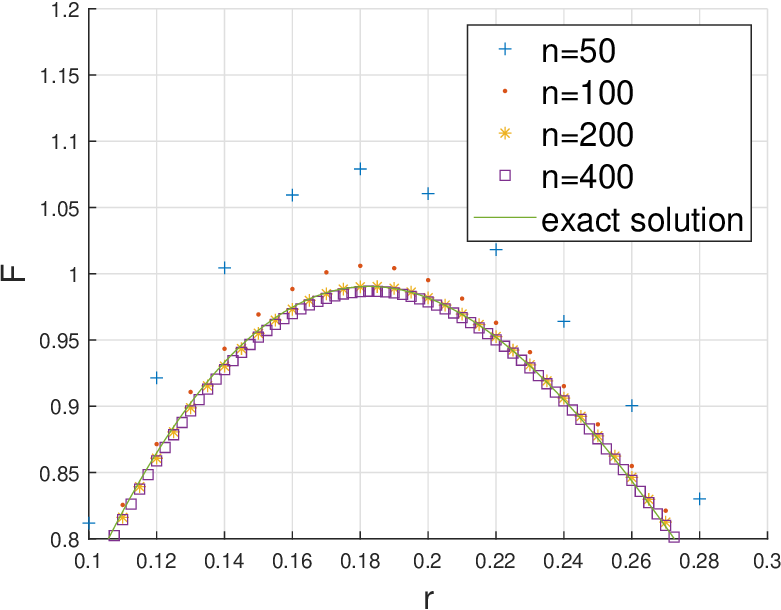}
  \caption{Results for $F$ using the MIRK1 method with $\kappa=10^2$ and CFL=$1/2^5$. Several spatial resolutions at $t=5$ are displayed. The whole spatial domain is shown at the left panel and a zoom near the maximum value for $F$ is displayed at the right panel.}
  \label{fig:DLFk100mirk1}
\end{figure}

We start testing the first-order MIRK method, MIRK1 from now on, setting $a=b=0$. Radial grid with $n=50, 100, 200, 400$ cells and a CFL factor of $1/2^l$ with $l=0,1,\ldots,6$ were used. In \figref{fig:DLEk100mirk1} and \figref{fig:DLFk100mirk1} we show the results for $E$ and $F$ at $t=5$, respectively, with CFL=$1/2^5$, together with the analytical solution.

\begin{figure}[htbp!]
  \centering
      \includegraphics[width=0.6\linewidth]{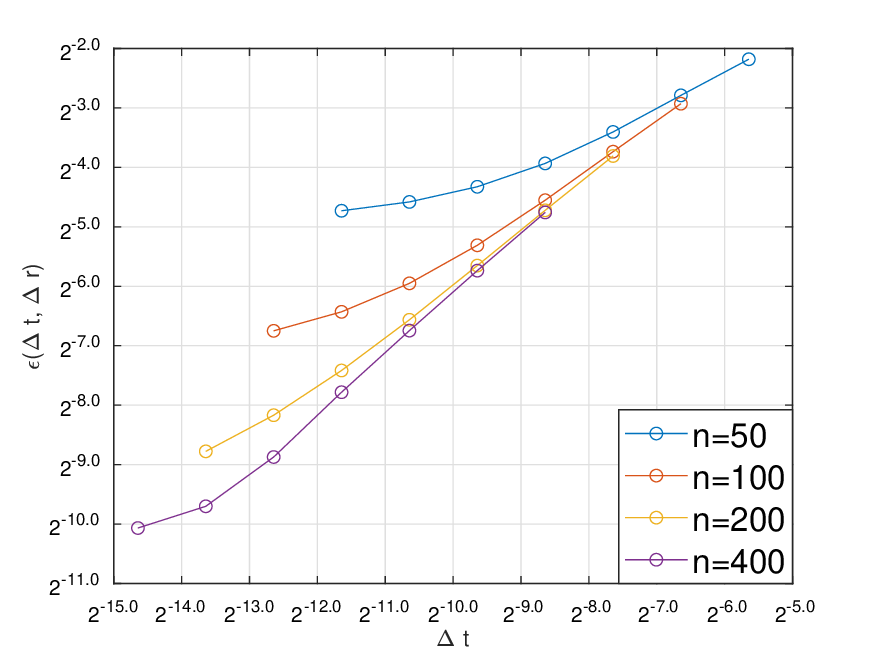}
  \caption{$L_2$ norm errors of $E$ between the exact and numerical solutions divided by $E(t,r=0)$ as function of $\Delta t$, using the MIRK1 method and $\kappa=10^2$. CFL=$1/2^l$ with $l=0,1,\ldots,6$ (values from right to left for each curve) have been used.}
  \label{fig:DLEk100mirk1orders}
\end{figure}

Let us compute estimates of the order of convergence. We denote $\varepsilon(\Delta t,\Delta r)$ the $L_2$ norm of the error of the numerical solution for $E$ at a temporal resolution $\Delta t$ and a spatial resolution $\Delta r$ w.r.t.~the exact solution $E$, divided by $E(t,r=0)$. The results for all temporal ($\Delta t = \sqrt{3} \, \text{CFL}\,\Delta r$) and spatial resolutions ($\Delta r= 1/n$) are shown in \figref{fig:DLEk100mirk1orders}.

We find two regimes with a specific dependence of the error on $\Delta t$: for a larger time step, $\varepsilon(\Delta t,\Delta r)$ is described by a power law, $\varepsilon(\Delta t,\Delta r) \propto (\Delta t)^{p_t}$, while the error becomes independent of $\Delta t$ for higher temporal resolutions.  We interpret this behaviour following the results of \citep{Rembiasz2017}. The total error is the sum of an error due to the spatial and another one due to the temporal discretisation. Each error contributes with an in principle different power law. For most combinations $(\Delta r, \Delta t)$, one of the two terms is dominant. For instance, if $\Delta r$ is fixed, the temporal error dominates for poor temporal resolutions. In that case, reducing the CFL number reduces the total error accordingly. This is the power-law section of the curves. Once $\Delta t$ becomes so small that the temporal error is below the spatial one, a change in CFL no longer has any influence on the total error: we are in the asymptotically flat part of the curves. The same considerations hold for the spatial error term and the grid resolution. Determining the spatial/temporal convergence order via measuring the power-law index has to be done in the regime where the total error is dominated by the corresponding spatial/temporal component.

Keeping the CFL factor fixed at a sufficiently low value to ensure that the total error is not contaminated by the temporal term, and varying the spatial resolution accordingly, we estimate the spatial order of convergence via the following formula:
\begin{equation}
    p\approx \log_2\left(\frac{\varepsilon(\Delta t,\Delta r)}{\varepsilon(\Delta t/2,\Delta r/2)}\right),\;\; \Delta t = \sqrt{3} \, \text{CFL}\,\Delta r.
    \label{eq:ord}
\end{equation}
The results are shown in the first row of \tableref{tab:order100}. A second order of convergence is observed, except for the last column that corresponds to the values where the temporal error starts to dominate.

\begin{table}[h]
\begin{center}
\begin{tabular}{c|ccc}
\hline\hline
$\Delta r$ & $0.02$ & $0.01$ & $0.005$ \\ \hline
 
  $p$ for MIRK1 &  2.01974 &  2.02764 &   1.28970 \\
   $p$ for MIRK2  & 2.01781 &  2.02307 &  1.30959 \\
  \hline\hline
\end{tabular}
\end{center}
\caption{Estimations of spatial order of convergence $p$, according to formula \eqref{eq:ord}, for $\kappa=10^2$, $\Delta t =\sqrt{3}\,\text{CFL}\,\Delta r$ and CFL=$1/2^6$.}
\label{tab:order100}
\end{table}

On the other hand, estimates for the purely temporal order of convergence, for a fixed sufficiently low $\Delta r$, are computed with the fitted slope from the errors $\varepsilon(\Delta t,\Delta r)$ for $\text{CFL}=1/2^l$ with $l=0,1,2$. The results are shown in the first row of \tableref{tab:ordert100}. In this case, an order of convergence close to 1 is observed.

\begin{table}[h]
\begin{center}
\begin{tabular}{c|cccc}
\hline\hline
$\Delta r$ & $0.02$ & $0.01$ & $0.005$ & $0.0025$ \\ \hline
  $p_t$ for MIRK1  & 0.5863 & 0.7963 & 0.9211 & 1.0144 \\
  $p_t$ for MIRK2 &  0.5961 & 0.8017 & 0.9237 & 1.0150 \\
  \hline\hline
\end{tabular}
\end{center}
\caption{Estimations of temporal order of convergence $p_t$ (fixing $\Delta r$) for $\kappa=10^2$. For each $\Delta r$ we obtain the estimates $p_t$ from the fits of $\varepsilon(\Delta t,\Delta r)$ with respect $\Delta t=\sqrt{3}\,\text{CFL}\,\Delta r$ with $\text{CFL}=1,\,0.5,\,0.25$.}
\label{tab:ordert100}
\end{table}

\begin{figure}[htbp!]
  \centering
      \includegraphics[width=0.49\linewidth]{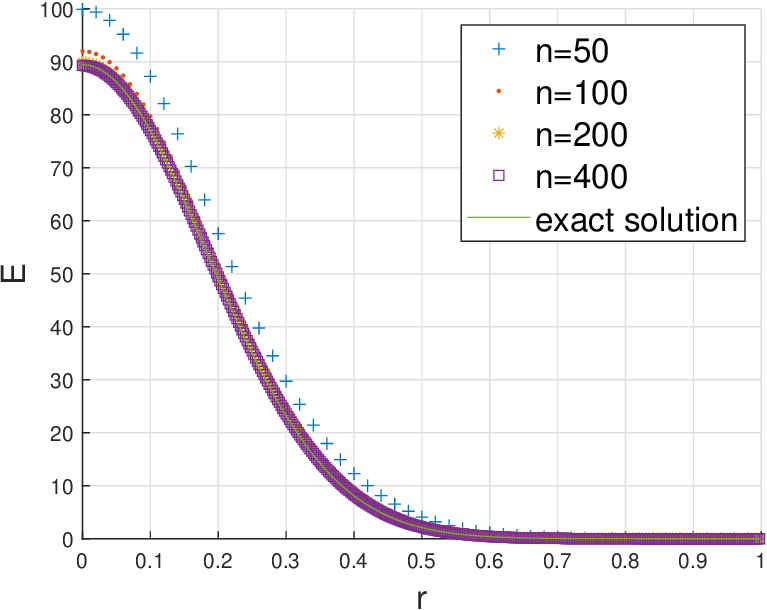}
      \includegraphics[width=0.49\linewidth]{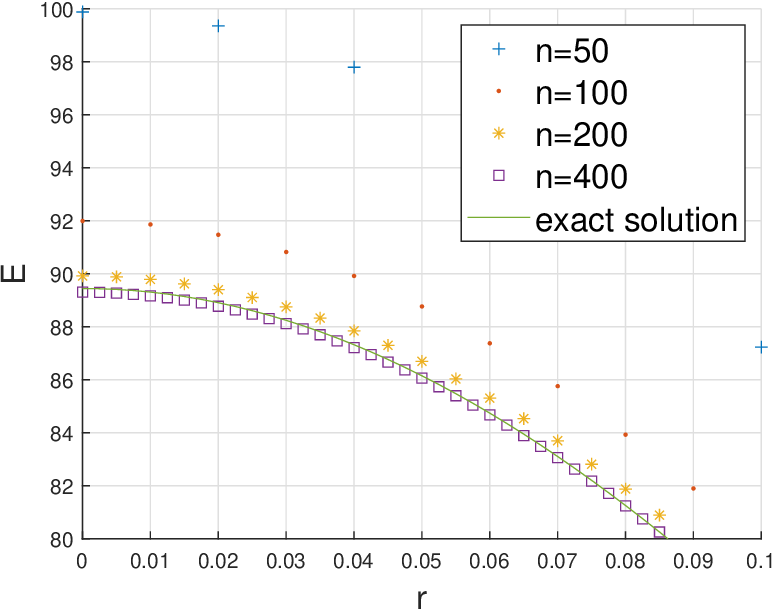}
  \caption{Results for $E$ using the MIRK2 method with $\kappa=10^2$ and CFL=$1/2^5$. Several spatial resolutions at $t=5$ are displayed. The whole spatial domain is shown at the left panel and a zoom near the origin is displayed at the right panel.}
  \label{fig:DLEk100mirk2}
\end{figure}

\begin{figure}[htbp!]
  \centering
      \includegraphics[width=0.49\linewidth]{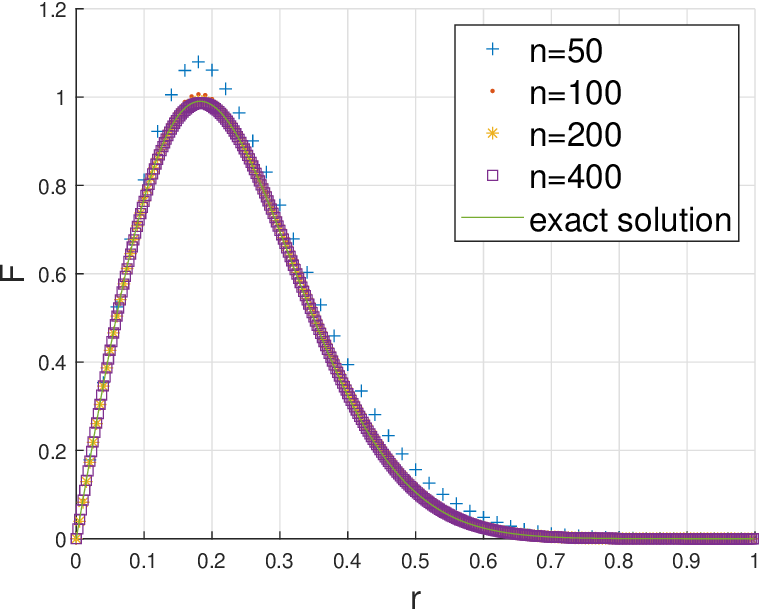}
      \includegraphics[width=0.49\linewidth]{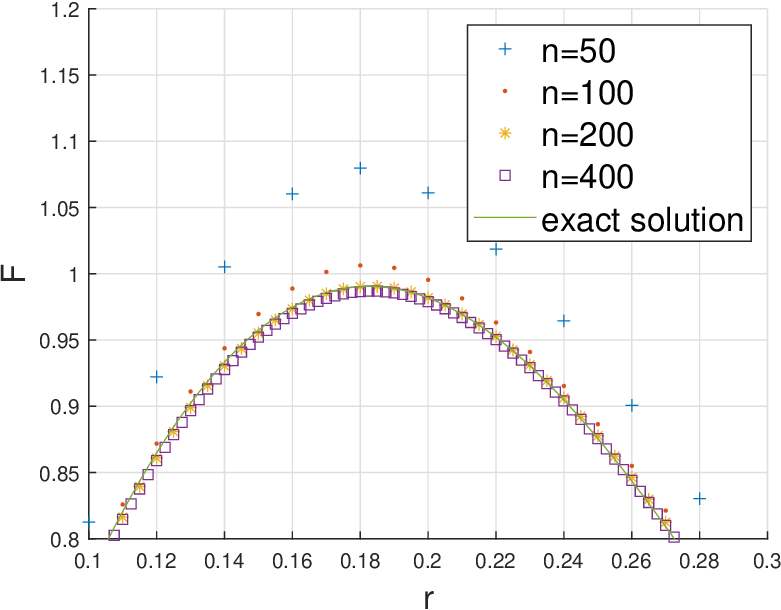}
  \caption{Results for $F$ using the MIRK2 method with $\kappa=10^2$ and CFL=$1/2^5$. Several spatial resolutions at $t=5$ are displayed. The whole spatial domain is shown at the left panel and a zoom near the maximum value for $F$ is displayed at the right panel.}
  \label{fig:DLFk100mirk2}
\end{figure}

We now test the second-order MIRK method, MIRK2 from now on, setting $a=b=1/2$, $a'=(a-1)/2$, $b'=(b-1)/2$. This election for the coefficients is based on the fact that for this test all the variables evolved are smooth, and we can try to recover second order of convergence also close to the stiff limit. The non-smoothness of the evolved variables in other numerical tests will force us to avoid these values for the coefficients in the MIRK method to get stable simulations. The same spatial resolutions and CFL factors as for MIRK1 are used. In \figref{fig:DLEk100mirk2} and \figref{fig:DLFk100mirk2} we show the results for $E$ and $F$ at $t=5$, respectively, with CFL=$1/2^5$, together with the analytical solution.

\begin{figure}[htbp!]
  \centering
      \includegraphics[width=0.6\linewidth]{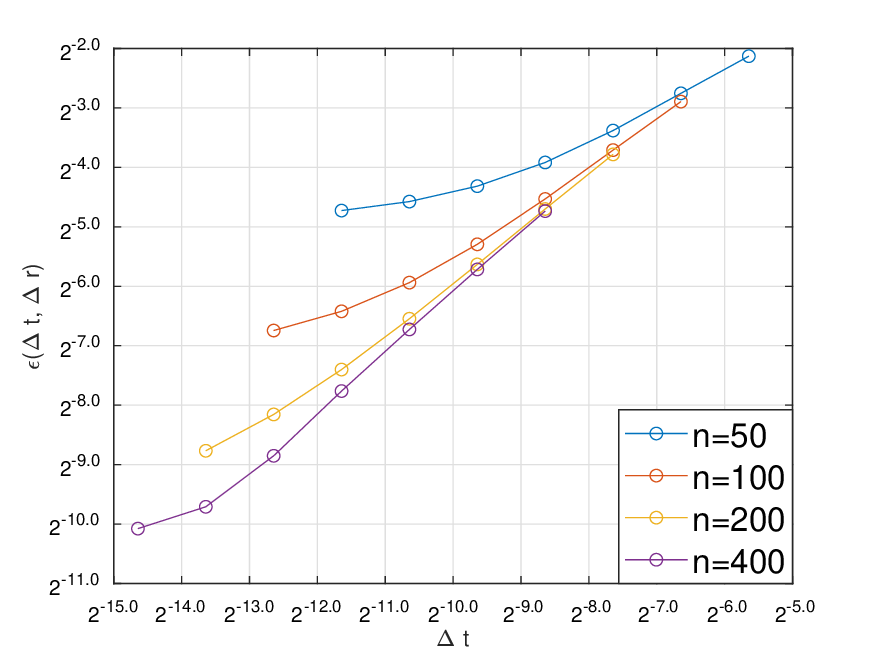}
  \caption{$L_2$ norm errors of $E$ between the exact and numerical solutions divided by $E(t,r=0)$ as function of $\Delta t$, using the MIRK2 method and $\kappa=10^2$. CFL=$1/2^l$ with $l=0,1,\ldots,6$ (values from right to left for each curve) have been used.}
  \label{fig:DLEk100mirk2orders}
\end{figure}

The errors $\varepsilon(\Delta t,\Delta r)$ for the same spatial and temporal resolutions as for the MIRK1 method are shown in \figref{fig:DLEk100mirk2orders}. A second order of convergence for the spatial error at fixed CFL factor (see the second row of \tableref{tab:order100}) is observed, before entering the regime in which the temporal error dominates over the spatial one. A first order of convergence for the temporal error is observed (see the second row of \tableref{tab:ordert100}). This reduction of the expected order of convergence to first order can be motivated by the use of a monotone method, as it is stated in \citep{leveque1992numerical}. This reduction has been also observed in several experiments using diagonally implicit Runge-Kutta methods (see e.g. \citep{Verwer1986-1975} and references thereby); as the MIRK methods share a similar strategy, the same reasons can lead to a reduction in the order of convergence.\\

\begin{figure}[htbp!]
  \centering
      \includegraphics[width=0.6\linewidth]{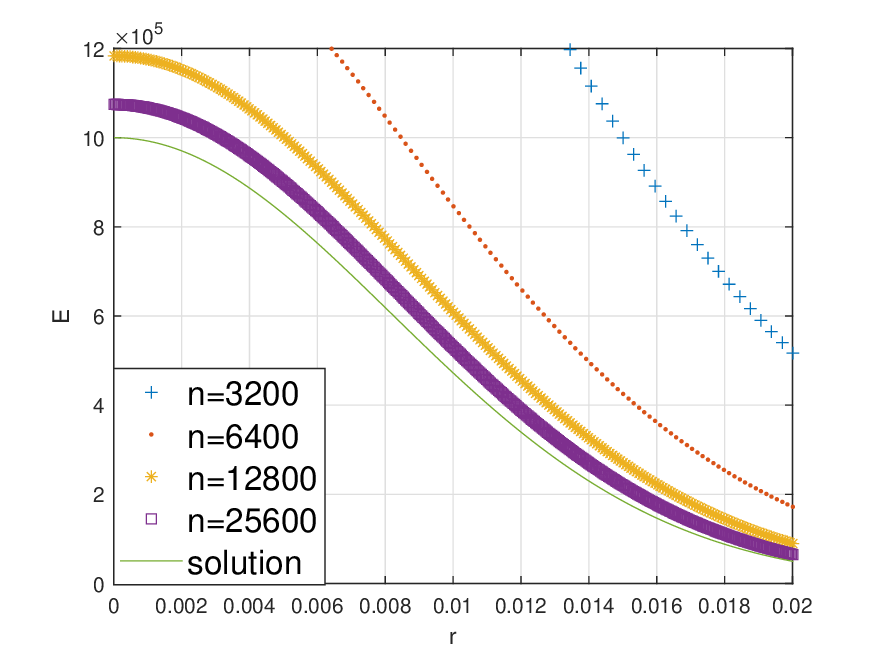}
  \caption{Results for $E$ using the MIRK1 method with $\kappa=10^5$ and CFL=$1/2^5$. Several spatial resolutions at $t=10$ are displayed.}
  \label{fig:DLEk1e5mirk1sl}
\end{figure}

We consider now a much higher value for the opacity, $\kappa=10^5$. We keep using a piece-wise linear reconstruction method for $S_{E,F}$, with a minmod slope-limiter and similar boundary conditions as for the $\kappa=10^2$ case. Now, we use spatial resolutions with $n=100\cdot 2^l$, $l=5,\ldots,8$. For each one, we consider CFL=$1/2^l$ with $l=0,1,\ldots,6$. We first start with the MIRK1 method setting $a=b=0$. In \figref{fig:DLEk1e5mirk1sl} we show the results for $E$ at $t=10$ with CFL=$1/2^5$, together with the analytical solution. Similar results are obtained for $F$.

\begin{figure}[htbp!]
  \centering
      \includegraphics[width=0.6\linewidth]{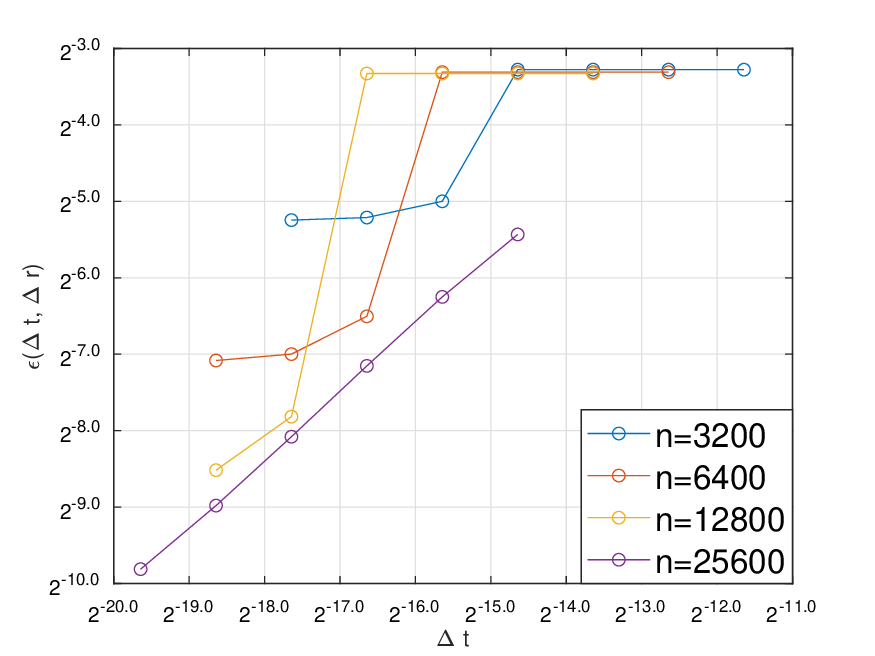}
  \caption{$L_2$ norm errors of $E$ between the exact and numerical solutions divided by $E(t,r=0)$ as function of $\Delta t$, using the MIRK1 method and $\kappa=10^5$. CFL=$1/2^l$ with $l=0,1,\ldots,5$ (values from right to left for each curve) have been used. The case $l=6$ for $n=3200,\, 6400$ has been also included.}
  \label{fig:DLk1e5mirk1slorders}
\end{figure}

The errors $\varepsilon(\Delta t, \Delta r)$ for several spatial and temporal resolutions are shown in \figref{fig:DLk1e5mirk1slorders}. For each $\Delta r$, $\Delta t=\sqrt{3}\,\text{CFL}\,\Delta r$ withCFL=$1/2^l$, $l=0,1,\ldots,5$, are used. Besides, we include $k=6$ for $n=3200,6400$. For $n=3200, 6400, 12800$, and CFL=$1/2^l$, $l=0,1,2,3$, we simply think that we are not still in the convergence regime. For $n=3200,6400$, and smaller time steps, we observe that for each value of $n$ a saturation (almost no dependence on $\Delta t$) of the error is rapidly achieved. One estimate for the spatial order of convergence can be computed for CFL=$1/2^6$ and using $n=3200,6400$, (simulations with $n=12800,25600$ are far to achieve a saturation for smaller time steps and cannot be used); a value close to second order of convergence is obtained (see first row of \tableref{tab:order1e5sl}). Furthermore, due to the previous reasons, we can only estimate the temporal order of convergence using $n=25600$, where the error depends on $\Delta t$ as expected. We calculate $p_t$ with a fit, as described previously, using the errors from simulations with CFL$=1/2^l$ and $l=0,1,...,5$; a value close to first order of convergence is obtained (see first row of \tableref{tab:order1e5sltmp}).

\begin{table}[h]
\begin{center}
\begin{tabular}{c|c}
\hline\hline
$\Delta r$ & $1/3200$ \\ \hline
  $p$ for MIRK1  & 1.83682   \\
  $p$ for MIRK2 &  1.83682  \\
  \hline\hline
\end{tabular}
\end{center}
\caption{Estimations of spatial order of convergence $p$, according to formula \eqref{eq:ord}, for $\kappa=10^5$, $\Delta t =\sqrt{3}\text{CFL}\Delta r$ and $\text{CFL}=1/2^6$.}
\label{tab:order1e5sl}
\end{table}

\begin{table}[h]
\begin{center}
\begin{tabular}{c|cc}
\hline\hline
$\Delta r$ & $1/12800$ & $1/25600$ \\ \hline
  $p_t$ for MIRK1  & - & 0.89246   \\
  $p_t$ for MIRK2 & 0.76945 & 0.88615 \\
  \hline\hline
\end{tabular}
\end{center}
\caption{Estimations of temporal order of convergence $p_t$ (fixing $\Delta r$) for $\kappa=10^5$. For each $\Delta r$ we obtain the estimates $p_t$ from the fits of $\varepsilon(\Delta t,\Delta r)$ with respect $\Delta t=\sqrt{3}\,\text{CFL}\,\Delta r$ with $\text{CFL}=1/2^{l}$ with $l=0,1,...,5$.}
\label{tab:order1e5sltmp}
\end{table}

\begin{figure}[htbp!]
  \centering
      \includegraphics[width=0.6\linewidth]{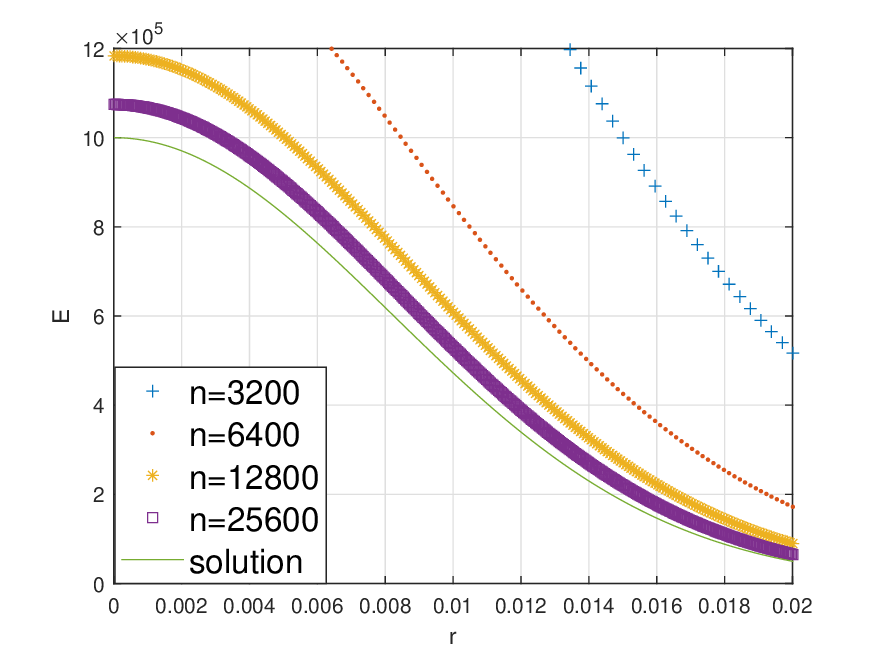}
  \caption{Results for $E$ using the MIRK2 method with $\kappa=10^5$ and CFL=$1/2^5$. Several spatial resolutions at $t=10$ are displayed.}
  \label{fig:DLEk1e5mirk2sl}
\end{figure}

\begin{figure}[htbp!]
  \centering
      \includegraphics[width=0.6\linewidth]{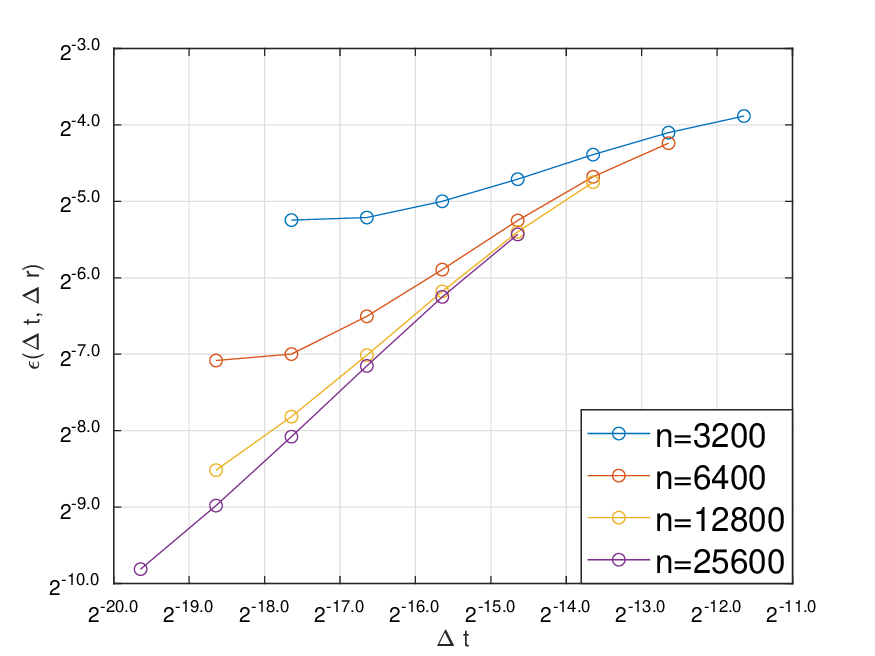}
  \caption{$L_2$ norm errors of $E$ between the exact and numerical solutions divided by $E(t,r=0)$ as function of $\Delta t$, using the MIRK2 method and $\kappa=10^5$. CFL=$1/2^l$ with $l=0,1,\ldots,5$ (values from right to left for each curve) have been used. The case $k=6$ for $n=3200,\, 6400$ has been also included.} 
  \label{fig:DLk1e5mirk2slorders}
\end{figure}

The results for $E$ at $t=10$ with CFL=$1/2^5$ using the MIRK2 method with $a=b=1/2$, $a'=(a-1)/2$, $b'=(b-1)/2$ are shown in \figref{fig:DLEk1e5mirk2sl}. Similar results are obtained for $F$. The corresponding errors are shown in \figref{fig:DLk1e5mirk2slorders}. The estimate of the spatial order of convergence is computed similarly as we did as for the MIRK1 case, using simulations with $n=3200, 6400$ and CFL=$1/2^6$; a value close to second order of convergence is obtained (see second row of \tableref{tab:order1e5sl}), as in the MIRK1 case. We have used the $n=12800, 25600$ simulations to estimate the temporal order of convergence; a value close to first order of convergence is obtained (see second row of \tableref{tab:order1e5sltmp}). We can justify this reduction in the temporal order of convergence as we have already done for $\kappa=10^2$.

An important difference between the results obtained for the MIRK1 and the MIRK2 methods is that in the case of the MIRK2 method all the considered resolutions lie within the convergence regime, while in the case of the MIRK1 method this only happens when higher resolutions are considered.

As a final comment, for both $\kappa=10^2$ and $\kappa=10^5$, it turns out that the errors are much smaller (several orders of magnitude) if second order centered differences instead of the linear piece-wise reconstruction scheme are used for the spatial derivatives, for the same temporal and spatial resolutions. This behaviour does not affect the estimates of the orders of convergence in general. For $\kappa=10^5$, some of the lower resolutions used in this section, and even lower resolutions, do lie within the convergence regime when second order centered differences are used, allowing to compute more estimates of the order of convergence (for both MIRK1 and MIRK2), but resulting in similar values.

\subsection{Toy model for a proto-neutron star}

\newcommand{\Ebb}{E_{\mathrm{eq}}}

\subsubsection{Setup}

One of the main simplifications of the test presented in \secref{sec:Pons1} is the absence of energy exchange between neutrinos and the static background. We propose in this subsection a new toy model for neutrino emission in a proto-neutron star (PNS) and its atmosphere including these processes. It is based on the solution of the spherically symmetric, grey (i.e., integrated over neutrino energies), $M_1$ equations of radiative transfer coupled to an equation for the evolution of the internal energy of the gas, $\epsilon$:
\begin{eqnarray}
    \label{eq:PNStoy-E}
    \partial_t E + \frac{1}{r^2} \partial_r (r^2 F) 
    & = & \kappa_{\mathrm{a}} (\Ebb - E),
    \\
    \label{eq:PNStoy-F}
    \partial_t F +  \frac{1}{r^2} \partial_r [r^2 P(E,F)]
    & = & - \kappa F,
    \\ 
    \label{eq:PNStoy-Ei}
    \partial_t \epsilon & = & - \kappa_{\mathrm{a}} ( \Ebb - E ).
\end{eqnarray}
All quantities are adimensional, $c=1$ is considered and $P(E,F)$ will be specified later. Neutrino-matter interactions are described by an absorption opacity, $\kappa_{\mathrm{a}}$, and a transport opacity, $\kappa = \kappa_{\mathrm{a}} + \kappa_{\mathrm{s}}$, that includes absorption and scattering processes. We do not account for the multi-flavour nature of the neutrinos and the associated exchange and transport of lepton number. The internal energy is related to the gas density $\rho$ and temperature $T$ via a simplified equation of state (EoS) that combines a polytropic, $\epsilon_{\mathrm{p}}$, and a thermal, $\epsilon_{\mathrm{t}}$, contribution, as follows:
\begin{equation}
    \epsilon = \epsilon_{\mathrm{p}} + \epsilon_{\mathrm{t}} = K_1 \rho^{\Gamma} + K_2 \rho T.
\end{equation}
We set $K_1 = 2$ and $\Gamma = 2$ and adjust $K_2$ such as to fix a ratio $\epsilon / \epsilon_{\mathrm{p}} = 2$ at $r = 0$; with this EoS and the initial density and temperature profiles introduced below, we get $K_2 \approx 10$.  The equilibrium energy density corresponding to a blackbody is approximated as $\Ebb = B_1 T^4 + B_2 \rho^4$, with $B_1 = 10^{-3}$ and $B_2 = 10^{-2}$. 

\begin{figure}
    \centering
      \includegraphics[width=0.6\linewidth]{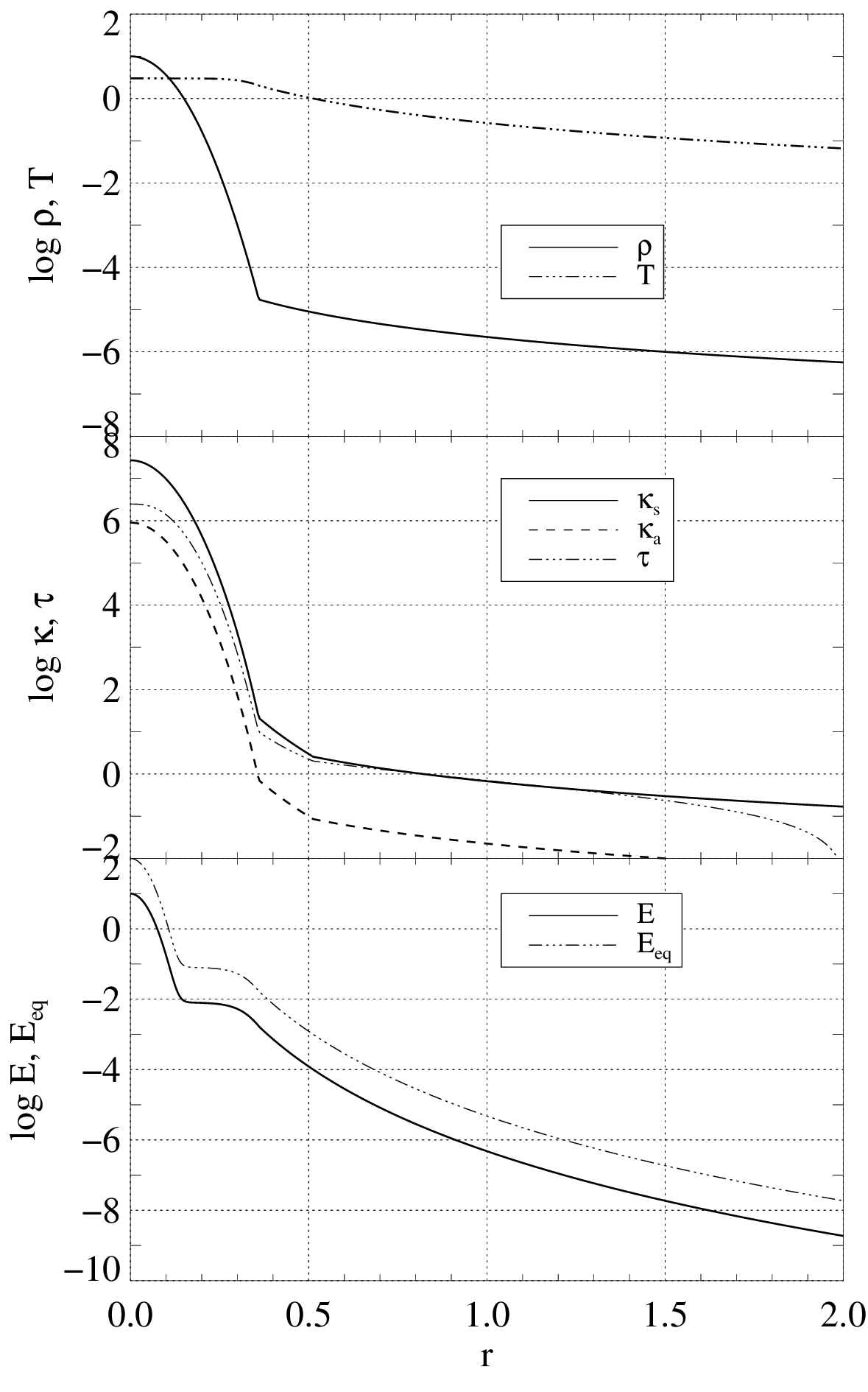}
    \caption{Initial data of the toy model for PNS cooling. Top panel: density and temperature profiles. Middle panel: profiles of the absorption and scattering opacity and optical depth. Bottom panel: blackbody equilibrium and initial neutrino energy density.}
    \label{fig:pnstoyinit}
\end{figure}

We set up the profile of (dimensionless) density as a combination of a parabolic function in the centre and a wind-like power law outside of it (see top panel of \figref{fig:pnstoyinit}, solid line):
\begin{equation}
\label{eq:PNStoy-rho}
    \rho (r) = \max [ \rho_{\mathrm{core}}(r), \, \rho_{\mathrm{atmo}}(r) ],
\end{equation}
where
\begin{eqnarray}
    \log_{10} [\rho_{\mathrm{core}}(r)] &=& \bar{\rho}_{\mathrm{core}} - (r/R_{\mathrm{core}})^2, \nonumber \\
    \rho_{\mathrm{atmo}}(r) &=&  10^{\bar{\rho}_{\mathrm{atmo}}} \cdot (r/R_{\mathrm{core}})^{-2}, \nonumber
\end{eqnarray}
$\bar{\rho}_{\mathrm{core}} = 1$, $\bar{\rho}_{\mathrm{atmo}} = -4$ and $R_{\mathrm{core}} = 0.15$. The atmosphere radius, $R_{\mathrm{atmo}}$, is the point where $\rho_{\mathrm{core}} (R_{\mathrm{atmo}}) = \rho_{\mathrm{atmo}} (R_{\mathrm{atmo}})$. As velocities are neglected, the density is kept constant. In contrast to the density, the gas temperature can vary during the simulation as energy is exchanged between matter and neutrinos. We initialise it according to
\begin{eqnarray}
    T (r) & = & 
    \left\{
    \begin{array}{cc}
    T_1 - \tanh\left(\frac{r - R_{\mathrm{atmo}}}{0.2 \, R_{\mathrm{atmo}}}\right),
    & r \le R_{\mathrm{atmo}}, \\
    T_{\mathrm{atmo}} \cdot (r/R_{\mathrm{atmo}})^{(-2)},
    & r > R_{\mathrm{atmo}},
    \end{array}
    \right. \hspace{0.5cm}
\end{eqnarray}
(top panel of \figref{fig:pnstoyinit}, dash-dot-dot-dotted line) where $T_1 = 1$, making the central temperature $T(r=0) = 3$; in addition, $T_{\mathrm{atmo}}$ follows from continuity at $R_{\mathrm{atmo}}$; initially, it is close to the central temperatures.

For the opacities, we use functions of density and temperature according to the approximations given by \citep{Janka2001} (see middle panel of \figref{fig:pnstoyinit}):
\begin{equation}
    \label{eq:PNStoy-kappa}
    \kappa_{\mathrm{a,s}} = k_{\mathrm{a,s}} \, \rho \, \max(T^2, T_{\kappa}^2),
\end{equation}
with $T_{\kappa} = 1$,  $k_{\mathrm{a}} = 10^4$ and $k_{\mathrm{s}} = 3 \cdot 10^5$. With these parameters, the configuration has a total optical depth of $\tau \gtrsim 10^6$. The transition between optically thick and optically thin regions, $\tau = 1$, lies in the atmosphere at $r \approx 0.8$.

The radiation is initialised with a low energy density, $E(t=0)=0.1\Ebb$, and no flux, $F(t=0)=0$ (see bottom panel of \figref{fig:pnstoyinit}).

\begin{figure}
    \centering
  \includegraphics[width=0.6\linewidth]{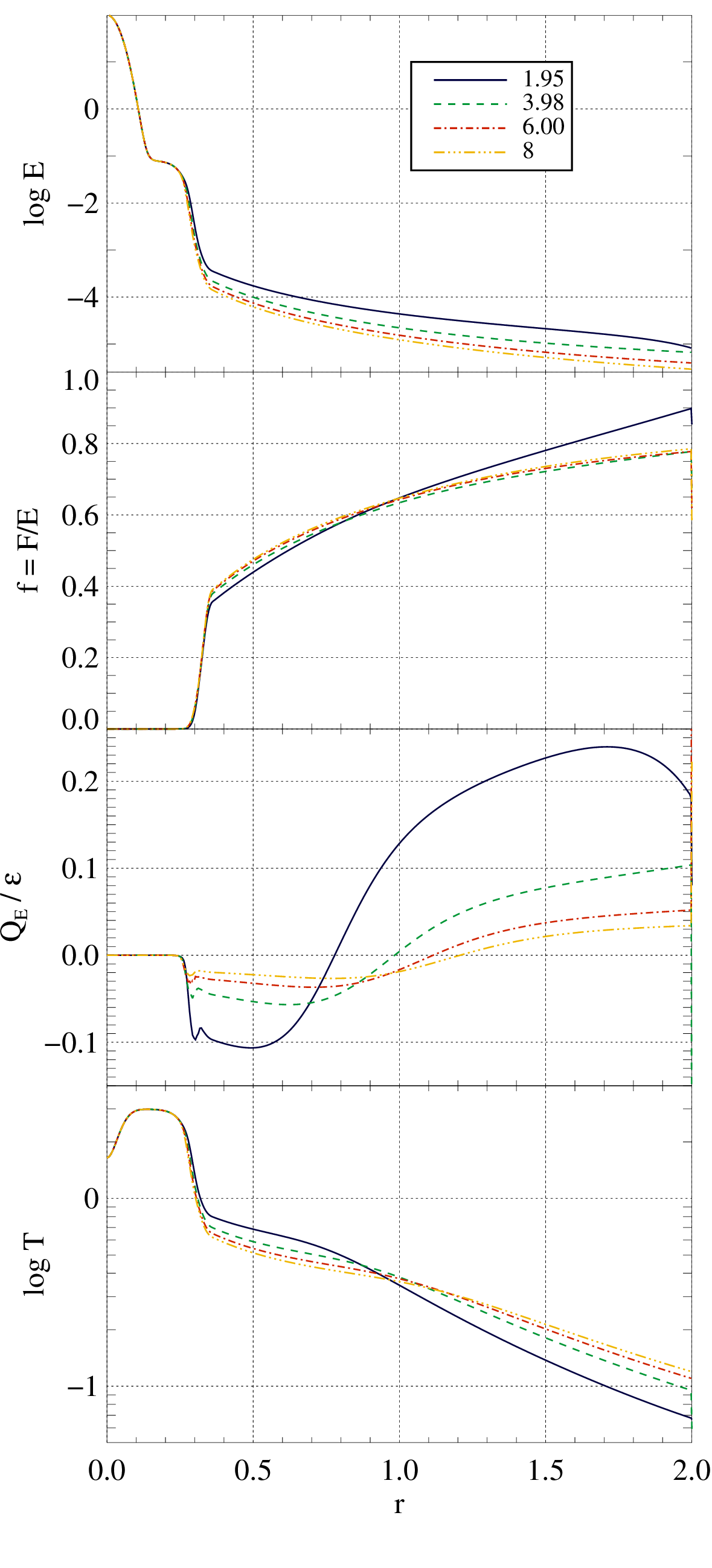}
    \caption{Radial profiles of the RK2 model of the PNS toy model at four times. The panels (top to bottom) show the logarithm of the neutrino energy density, $\log E$, the flux factor $f = F/E$, the rate at which the gas is heated normalised by the internal energy, $Q_E/\epsilon$, and the logarithm of gas temperature, $\log T$.}
    \label{fig:pnstoy-RK2-prof}
\end{figure}

We select the Minerbo closure \citep{minerbo1978} that relates the radiation pressure $P$ to the energy density as $P = \chi E$, with the Eddington factor $ \chi = (5 + 6 f^2 - 2 f^3 + 6 f^4)/15 $ depending on the flux factor $f = |F|/E$.

The system \eqref{eq:PNStoy-E}--(\ref{eq:PNStoy-Ei}) is solved on a uniform radial grid with $n_r = 200$ cells between $r = 0$ and $r = 1$ using a finite-volume discretisation and a second-order total variation diminishing (TVD) approach, in order to avoid undesirable oscillations in the numerical solution. Specifically, we use piecewise-linear reconstruction, as in the previous tests, but this time with the van Leer slope limiter. More information on TVD methods and the van Leer slope limiter can be found in \citep{leveque1992numerical}. The boundary conditions at the centre and the outer boundary are reflecting and outflow, respectively. We tested the same set of time integrators as in \secref{sec:Pons1}. Of those, we will compare here the results for a reference solution (see next subsection for more details) obtained with a second order Runge-Kutta and the second order MIRK method.

\subsubsection{Reference simulation: evolution with RK2}

We describe the evolution of the reference simulation performed with a second order Runge-Kutta time integrator using the first order MIRK method (with $a=b=0$) for the neutrino-matter interaction terms in each of the two substeps, RK2 from now on. This is the same method we regularly use in core-collapse simulations and thus serves as our reference. In the absence of other energy sources such as nuclear reactions or accretion, the only process is the gradual cooling of the hot core core by neutrino emission.

\begin{figure}
    \centering    \includegraphics[width=0.6\linewidth]{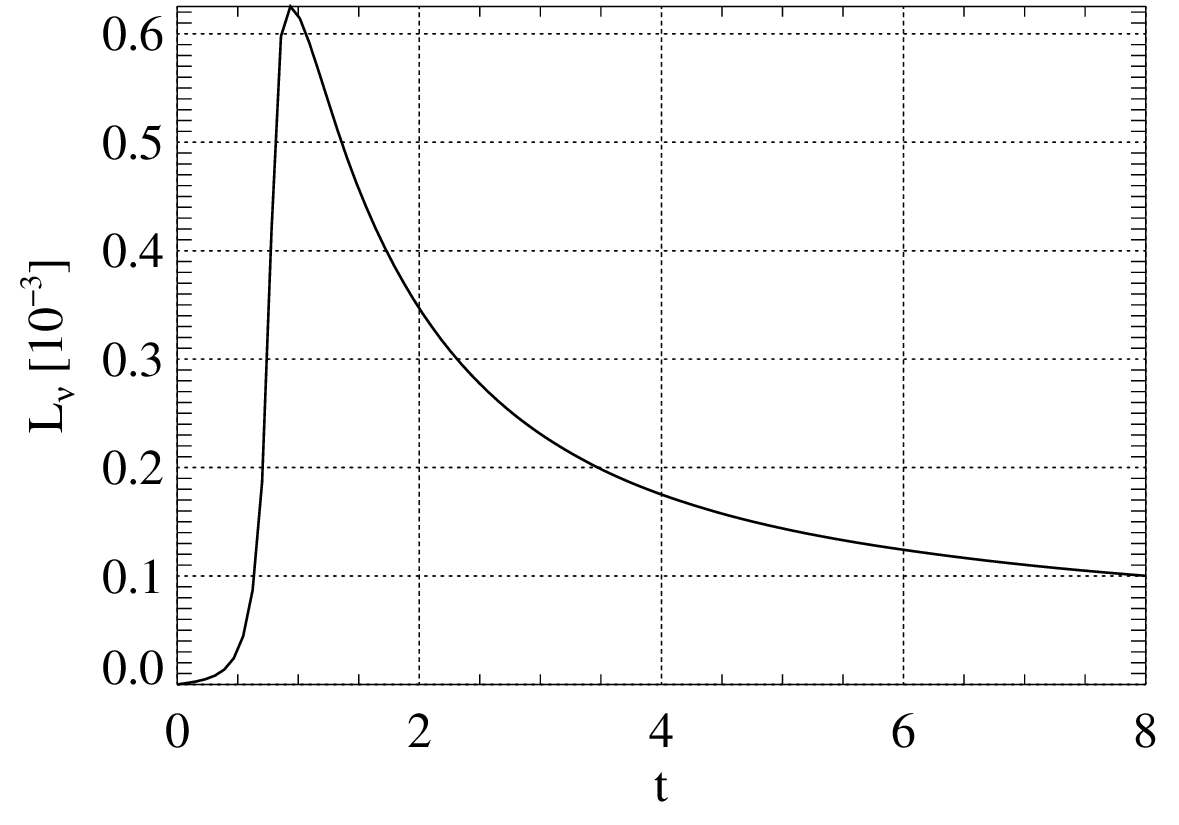}
    \caption{Time evolution of the neutrino luminosity of the PNS toy model simulated with the RK2 time integrator.}
    \label{fig:pnstoy-RK2-lum}
\end{figure}

We show profiles of the neutrino energy density and flux factor, the neutrino-matter energy exchange rate, and the matter temperature for four times up to the final time $t_{\mathrm{f}} = 8$ in \figref{fig:pnstoy-RK2-prof}, and the time evolution of the neutrino luminosity evaluated at $r \approx 1$ in \figref{fig:pnstoy-RK2-lum}.

The state starts to deviate from the initial conditions already very early. After a transient phase in which the artificial initial data for the neutrinos relax, the evolution slows down. Inside the PNS, the neutrino energy density approaches the (slowly evolving) blackbody distribution. The diffusion out of the dense core is slow with flux factors close to 0. The transition to free streaming in the atmosphere is quite abrupt with the flux factors jumping to $f \gtrsim 0.4$ over a short radial distance. Further out, $f$ tends to the asymptotic value of unity. In the atmosphere, we observe the formation of a gain radius separating neutrino cooling ($Q_E < 0$, decreasing gas temperature) above the PNS from neutrino heating ($Q_E > 0$, increasing temperature) at higher radii. The evolved temperature profile differs from the initial one in the off-centre maximum, which resembles that of detailed supernova simulations. The neutrino luminosity, $L_{\nu}$, has an early peak associated with the initial transient; afterwards, it decreases slowly as the PNS loses energy.

\begin{figure}[htbp!]
    \centering
      \includegraphics[width=0.6\linewidth]{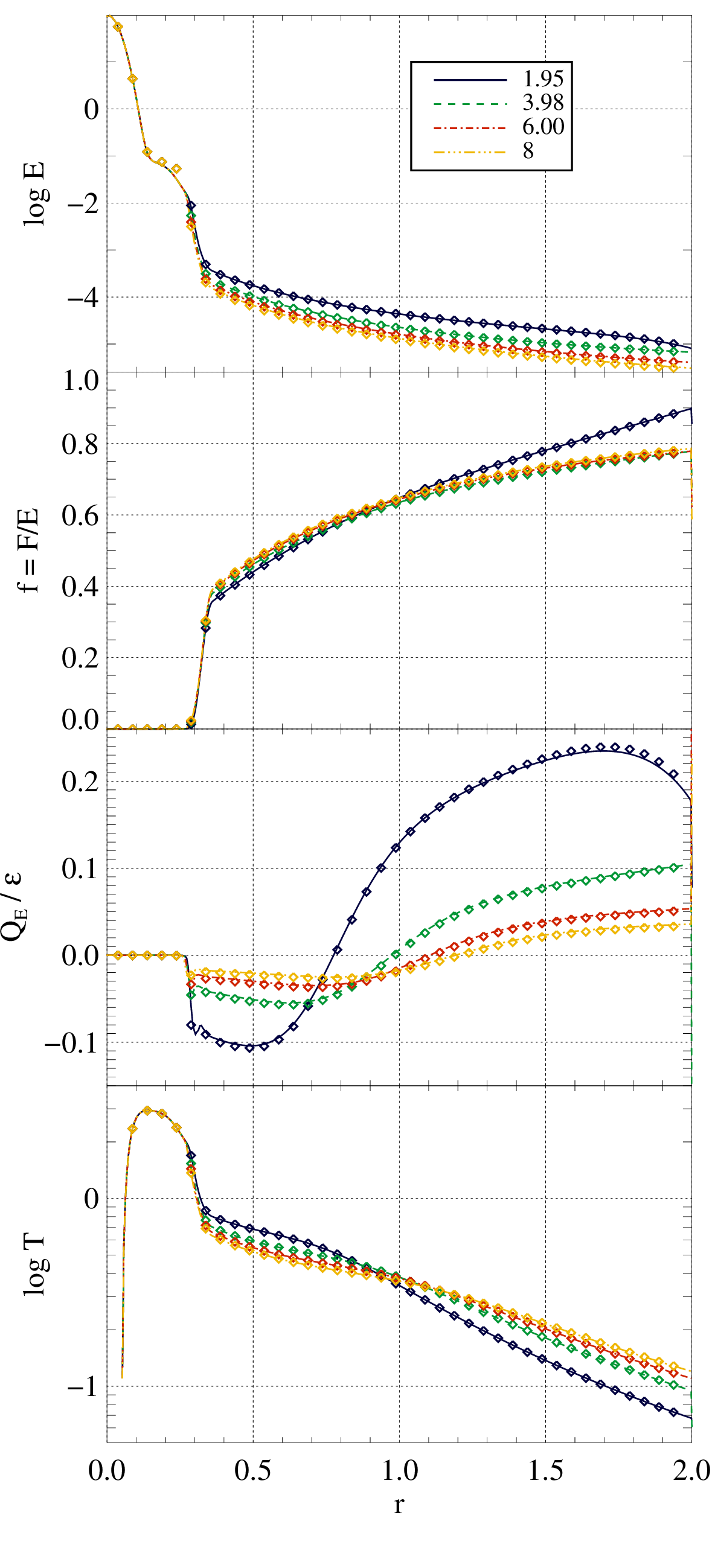}
    \caption{Same as \figref{fig:pnstoy-RK2-prof}.  The lines show the results of simulation MIRK2-1, symbols of the reference run with RK2 time integration.}
    \label{fig:pnstoy-MIRK2-prof}
\end{figure}

\subsubsection{MIRK2}

\begin{figure}[htbp!]
    \centering
      \includegraphics[width=0.6\linewidth]{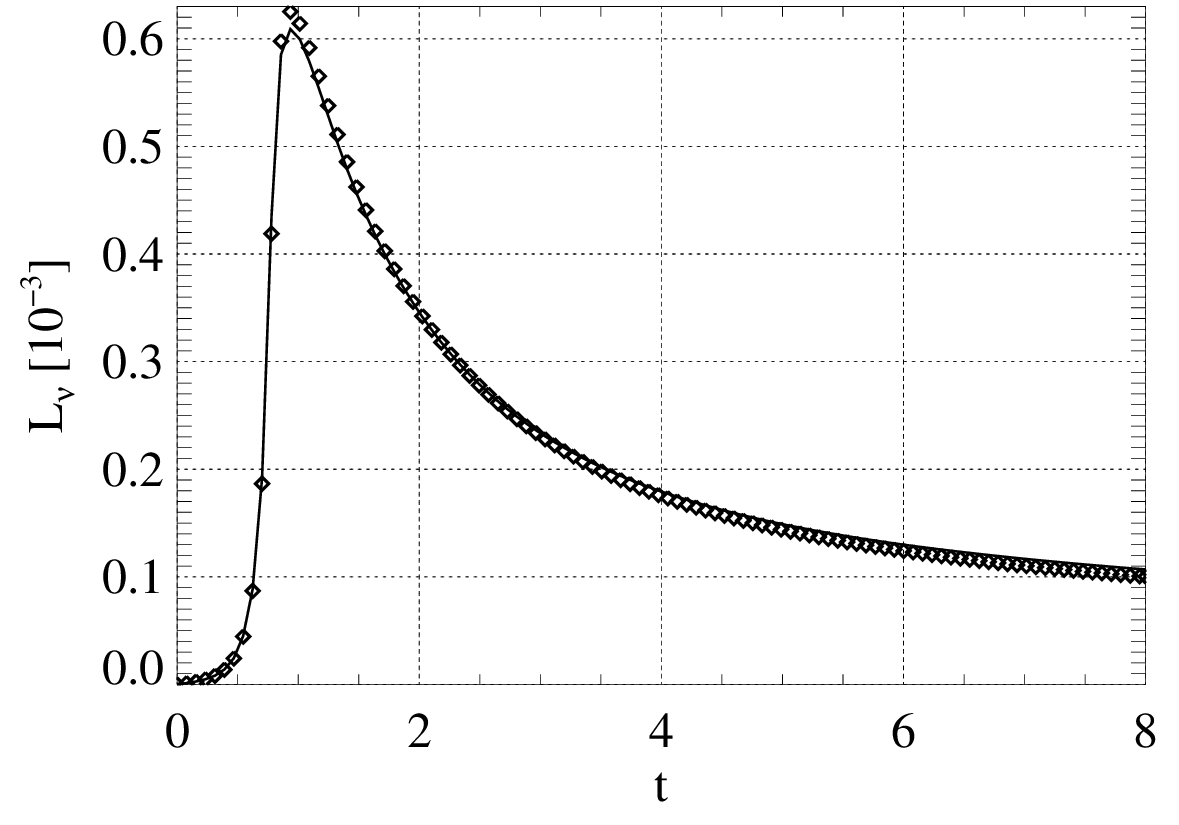}
    \caption{Same as \figref{fig:pnstoy-RK2-lum}, but comparing run MIRK2-1 run (lines) to the RK2 reference run (symbols).}
    \label{fig:pnstoy-MIRK2-lum}
\end{figure}

We show results of simulation MIRK2-1, with MIRK2 time integration and the parameter set
\begin{gather*}
    (a,a') = \left( a,\frac{(1-a)^2}{2a} \right) = (-1/4,-25/8), \\
    (b,b')=\left( b,\frac{(1-b)^2}{2b} \right),
\end{gather*} which, according to our analysis, should lie in the region of stability when non-smooth quantities are evolved, in \figref{fig:pnstoy-MIRK2-prof} and \figref{fig:pnstoy-MIRK2-lum}, together with the reference solution. As expected, the simulation runs stably and produces results very similar to the ones of the RK2 reference run.

\begin{figure}[htbp!]
    \centering
      \includegraphics[width=0.6\linewidth]{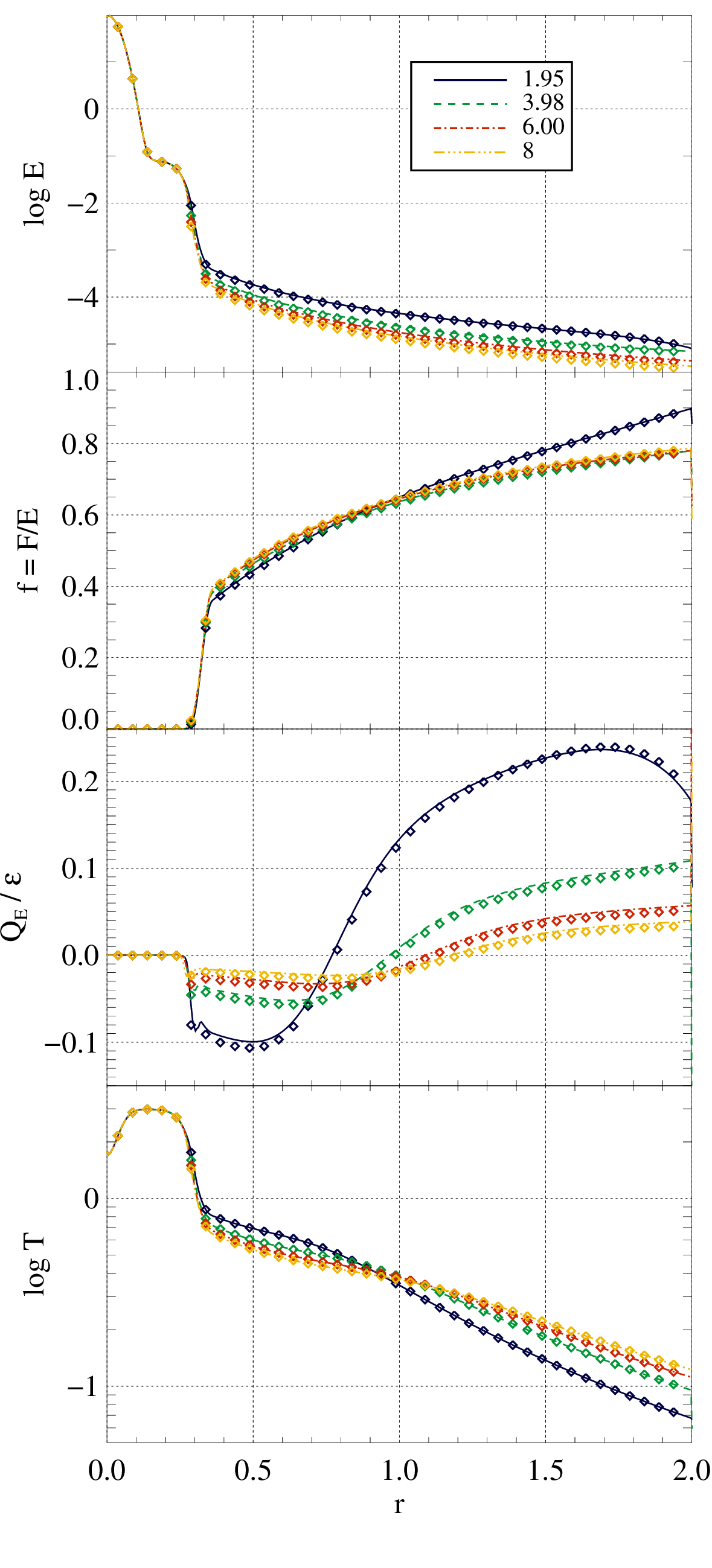}
    \caption{Same as \figref{fig:pnstoy-RK2-prof}. The lines show the results of simulation MIRK2-a, symbols of the reference run with RK2 time integration.}
    \label{fig:pnstoy-MIRK2a-prof}
\end{figure}

Having verified that nominally stable MIRK parameters indeed yield a stable simulation, we explore two variations: MIRK2-a  with $(a,a',b,b') = (-1/4,-5/8,-1/4,-25/8)$ (i.e., $b' = (1-b)^2/2b$ and $a' = (a-1)/2$) and MIRK2-b with $(a,a',b,b') = (-1/4,-25/8,-1/4,-5/8)$ (i.e., $b' = (b-1)/2$ and $a' = (1-a)^2/2a$). The former set of parameters produces results that are very similar to the reference and MIRK2-1 runs (see \figref{fig:pnstoy-MIRK2a-prof}), while the latter fails due to a numerical instability whose origin is at the surface of the core, i.e., in the region where the radiation decouples from the matter and the flux factor increases from $f \approx 0$ to $f \approx 0.4$. We interpret this finding in the sense that the neutrino momentum equation is more sensitive to the correct choice of parameters than the energy equation. This finding will be explored in greater detail below in the context of a core-collapse simulation.

\subsection{Core-collapse supernovae test}
\subsubsection{Input physics}
\label{sec:inphys}

To assess its properties, we apply our method to the radiation-hydrodynamics of stellar core-collapse in spherical symmetry. This setup tests the scheme in a highly dynamic system including both the optically thin and the optically thick, stiff regimes of neutrino-matter interactions. As such, it represents a demanding problem for numerical codes. While the neglect of non-spherical flows limits the degree of realism, it makes the problem more standardised and controllable. Therefore, our tests follow in the footsteps of many previous studies of new schemes that used similar setups (e.g., \citep{just2015, Rampp_Janka__2002__AA__Vertex, Liebendorfer_et_al__2004__apjs__FD_Neutrino_GRRHD_Agile-Boltztran, Sekiguchi__2010__ProgressofTheoreticalPhysics__StellarCoreCollapseinFullGeneralRelativitywithMicrophysics--FormulationandSphericalCollapseTest, OConnor__2010__CQGra__A_new_open-source_code_for_spherically_symmetric_stellar_collapse_to_neutron_stars_and_black_holes, Mueller_et_al__2010__apjs__A_New_Multi-dimensional_General_Relativistic_Neutrino_Hydrodynamic_Code_for_Core-collapse_Supernovae.I.Method_and_Code_Tests_in_Spherical_Symmetry, OConnor__2015__TheAstrophysicalJournalSupplementSeries__AnOpenSourceNeutrinoRadiationHydrodynamicsCodeforCoreCollapseSupernovae, Kuroda_et_al__2016__apjs__ANewMulti-energyNeutrinoRadiation-HydrodynamicsCodeinFullGeneralRelativityandItsApplicationtotheGravitationalCollapseofMassiveStars,
Perego_et_al__2016__apjs__AnAdvancedLeakageSchemeforNeutrinoTreatmentinAstrophysicalSimulations, OConnor__2018__JournalofPhysicsGNuclearPhysics__GlobalComparisonofCoreCollapseSupernovaSimulationsinSphericalSymmetry, Just__2018__MonthlyNoticesoftheRoyalAstronomicalSociety__CoreCollapseSupernovaSimulationsinOneandTwoDimensionsComparisonofCodesandApproximations, Laiu__2021__TheAstrophysicalJournalSupplementSeries__ADGIMEXMethodforTwoMomentNeutrinoTransportNonlinearSolversforNeutrinoMatterCoupling}).

All simulations presented in the remainder of this section use the neutrino-(magneto-)hydrodynamics code Alcar \citep{just2015} and, except where explicitly stated, the same input physics, initial conditions, and, except for the time integration, numerical methods and parameters. We solve the equations of special relativistic hydrodynamics including a balance law for the electron fraction of the gas, $Y_e$. We account for the self-gravity of the star using a pseudo-relativistic gravitational potential (potential $A$ of \citep{Marek_etal__2006__AA__TOV-potential}). The spectral $M_1$ transport modules evolve the radiation energy and momentum density in a reference frame comoving with the fluid. The coupling between neutrino particle energies via velocity and gravitational terms, e.g., Doppler or gravitational red-/blue-shifts, are included up to first order in $v/c$. We describe the thermodynamic properties of the gas using the nuclear equation of state (EoS) SFHo \citep{Steiner_et_al__2013__apj__Core-collapseSupernovaEquationsofStateBasedonNeutronStarObservations}. Strictly speaking, an EoS of this type, assuming that the composition of the gas is given by nuclear statistical equilibrium, is not valid for low temperatures and densities. Nonetheless, we simplify our setup by not including a transition to a sub-nuclear EoS below a threshold density. This choice has no implication for the tests at hand because the neutrino-matter interaction rates are very small at the densities where the transition between EoS regimes would take place.

We employ the spectral $M_1$ transport methods for three species of neutrinos: electron neutrinos, $\nu_e$, and antineutrinos, $\bar{\nu}_e$, and one species, $\nu_X$, including the mu and tau neutrinos and their antiparticles. Our set of neutrino-matter reactions contains the important interactions that dominate the dynamics of core-collapse (see \citep{just2015, Just__2018__MonthlyNoticesoftheRoyalAstronomicalSociety__CoreCollapseSupernovaSimulationsinOneandTwoDimensionsComparisonofCodesandApproximations} for implementation details):
\begin{itemize}
\item absorption and emission of $\nu_e$ and $\bar{\nu_e}$ by $\beta$ processes of free neutrons and protons and nuclei,
\item iso-energetic scattering of neutrinos of all flavours off nucleons and nuclei,
\item pair creation of neutrinos of all flavours by electron-positron annihilation and nucleonic bremsstrahlung,
\item non-iso-energetic scattering of neutrinos of all flavours off electrons and positrons.
\end{itemize}
We note that the last process is not written in terms of an opacity and thus our MIRK method does not apply. We treat it in an operator-split manner in the same way as described in \citep{Just__2018__MonthlyNoticesoftheRoyalAstronomicalSociety__CoreCollapseSupernovaSimulationsinOneandTwoDimensionsComparisonofCodesandApproximations}. In principle, the same holds for the pair processes. However, the approximate treatment of \citep{OConnor__2015__TheAstrophysicalJournalSupplementSeries__AnOpenSourceNeutrinoRadiationHydrodynamicsCodeforCoreCollapseSupernovae} reformulates the interaction in terms of opacities, which allows us to include them in the MIRK scheme.

\subsubsection{Initial data and reference simulation}
\label{sSek:RefSim}

As a test case, we used the same model as in the comparison of neutrino-hydrodynamics codes of \citep{Rampp_Janka__2002__AA__Vertex}, i.e., the core of a star of a zero-age main-sequence mass of $M_{\mathrm{ZAMS}} = 15 \, \msol$. Before presenting results of the new MIRK method implementation, we describe the dynamics of a reference simulation, RK2 from now on, computed with the traditional scheme used in the Alcar code \citep{just2015}. It uses a method similar to our first order MIRK scheme (with $a=b=0$) as a building block in a second order Runge-Kutta time integrator.  

\begin{figure}[htbp!]
  \centering
      \includegraphics[width=0.6\linewidth]{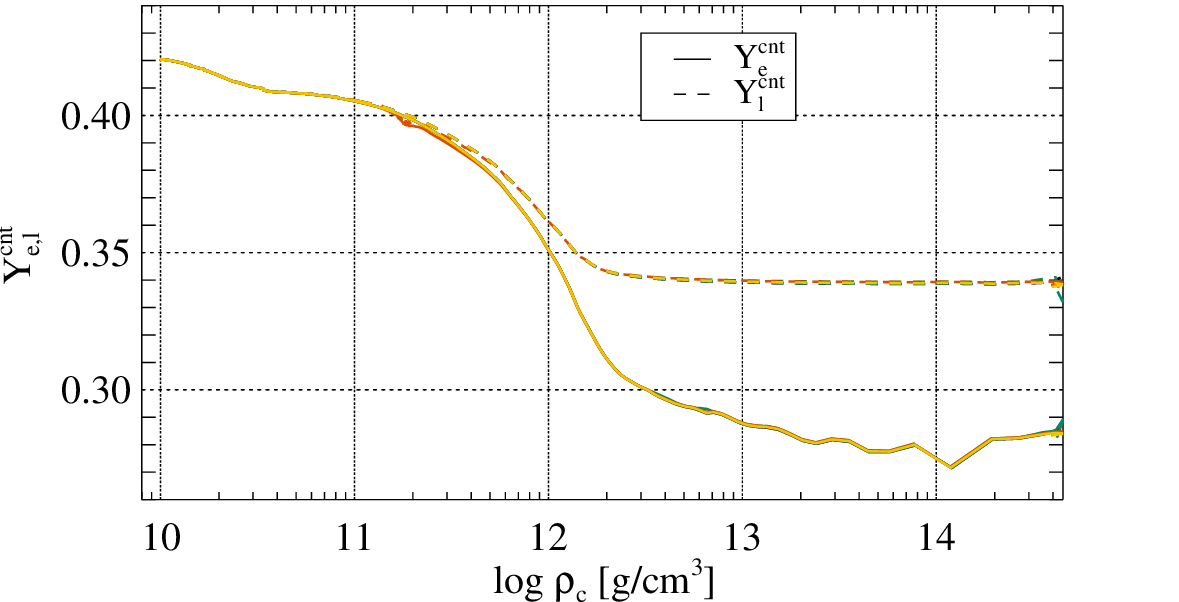}
      \includegraphics[width=0.6\linewidth]{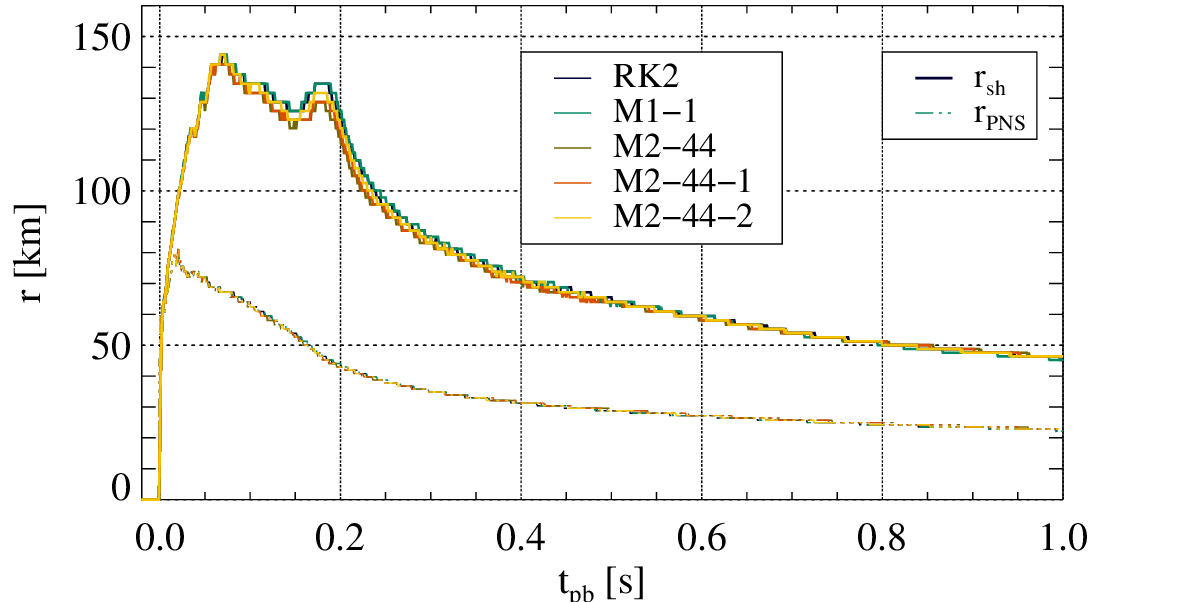}
      \includegraphics[width=0.6\linewidth]{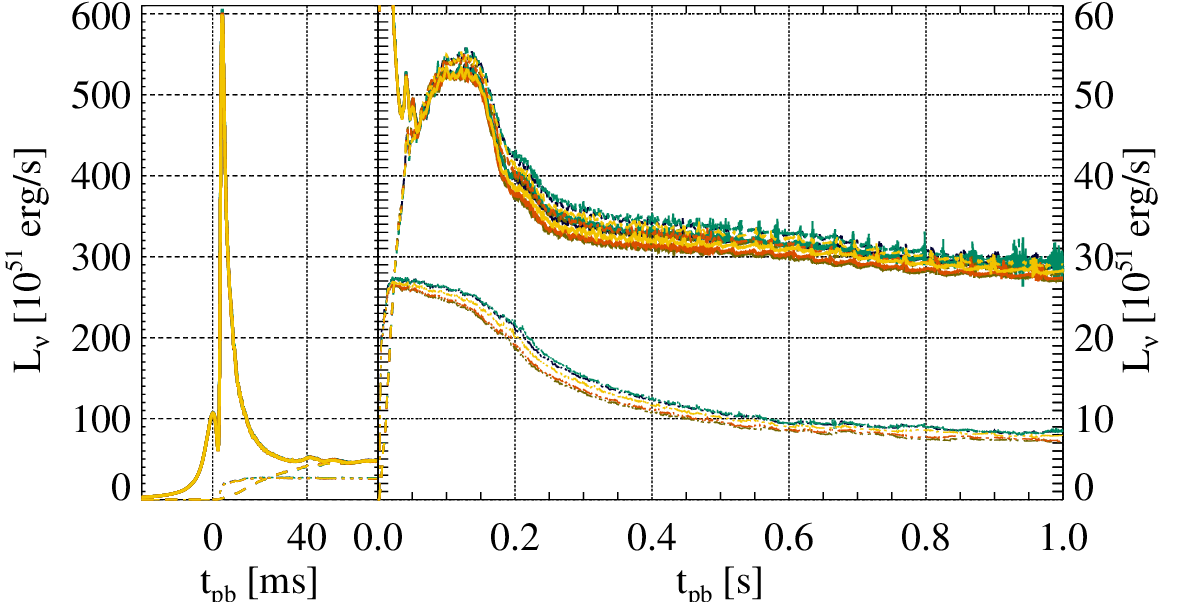}
      \includegraphics[width=0.6\linewidth]{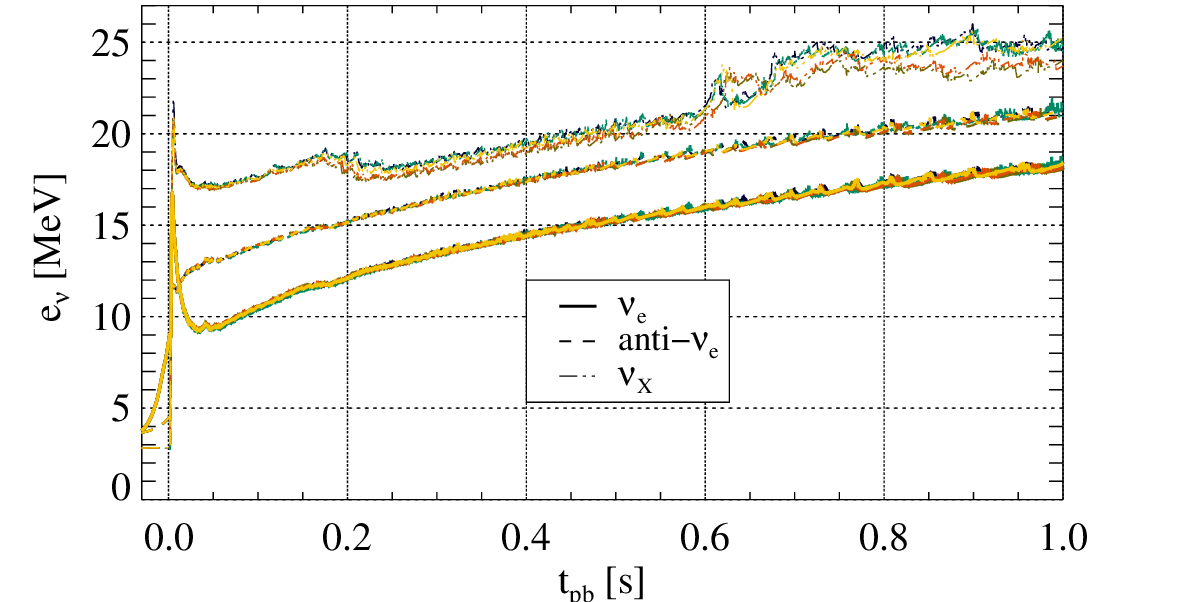}
  \caption{
    Evolution of important variables of our simulations. Models are distinguished by line colors, as indicated in the second panel.
    Top panel: Central electron and lepton fractions as a function of central density during collapse.
    Second panel: Time evolution of the radii of the shock, $r_{\mathrm{sh}}$, and the PNS, $r_{\mathrm{PNS}}$, during the first
    second after bounce. 
    Third panel: Luminosities of the three neutrino flavors (see legend in the bottom panel) as functions of time. The left part focuses on the $\nu_{e}$ burst, the right part shows the evolution until the end of the simulations.
    Bottom panel: Mean energy of the three neutrino energies as functions of time.
  }
  \label{Fig:rsh}
\end{figure}

As the central density increases during collapse, electron captures deleptonize the matter and drive the electron fraction at the center to values $Y_e^{\mathrm{cnt}} \approx 0.28$ and the lepton fraction, including the net lepton number corresponding to $\nu_e$ and $\bar{\nu_e}$, to $Y_l^{\mathrm{cnt}} \approx 0.34$. These quantities assume a roughly constant level once the neutrinos are trapped inside the inner core as densities $\rho \gtrsim 10^{12} \, \gccm$ render the gas optically thick. The shock wave launched at core bounce (time $t_{\mathrm{b}}$) and the formation of the proto-neutron star (PNS) stalls about 70 ms later after having reached a maximum radius of $r_{\mathrm{sh;max}} \approx 145 \, \mathrm{km}$, i.e., still inside the collapsing iron core (see \figref{Fig:rsh}, second panel). It recedes slowly for another 90 ms to $r_{\mathrm{sh}} \approx 125 \, \mathrm{km}$. Matter continues to fall through the shock wave and settles onto the PNS, which gradually contracts from a maximum radius of up to $r_{\mathrm{PNS}} \lesssim 80 \, \mathrm{km}$ immediately after bounce to $r_{\mathrm{PNS}} \gtrsim 20 \, \mathrm{km}$ at $t_{\mathrm{pb}} = 1 \, \mathrm{s}$ (here we use the radius of the $\nu_e$-sphere as a proxy for the PNS radius). By $t_{\mathrm{pb}} =
t - t_{\mathrm{b}} \approx 150 \, \mathrm{ms}$, the entire iron core has been accreted. Consequently, the density and ram pressure of the accreting matter drops, which causes a brief expansion of the shock by about 10 km. Neutrinos heat the post-shock gas, but, as it is typical for spherically symmetric core collapse, the conditions for shock revival and an explosion are never met. Thus, the shock wave gradually contracts to a radius below 50 km over the course of 1s after bounce. The early neutrino emission is characterized by the intense burst of $\nu_e$ emitted in the first few tens of ms after bounce (third panel of \figref{Fig:rsh}). After the burst, the $\nu_e$ luminosity, $L_{\nu_e}$, and those of the other two flavors reach slowly varying values of several $10^{52} \, \mathrm{erg}/\mathrm{s}$. We find the typical ordering with almost equal luminosities of the electronic flavors and a lower emission of the heavy-lepton neutrinos as well as the dependence of the luminosities on the mass accretion rate that leads to the lower levels of $L_{\nu}$ after $t_{\mathrm{pb}} \sim 200 \, \mathrm{ms}$. The mean energies (bottom panel) with values in the range of 10--25 MeV reflect the rising temperatures near three neutrinospheres of the three flavors with the sequence $e_{\nu_e} < e_{\bar{\nu_e}} < e_{\nu_{X}}$ following from the hierarchy of neutrino-matter cross sections.

\begin{figure}[htbp!]
  \centering
      \includegraphics[width=0.6\linewidth]{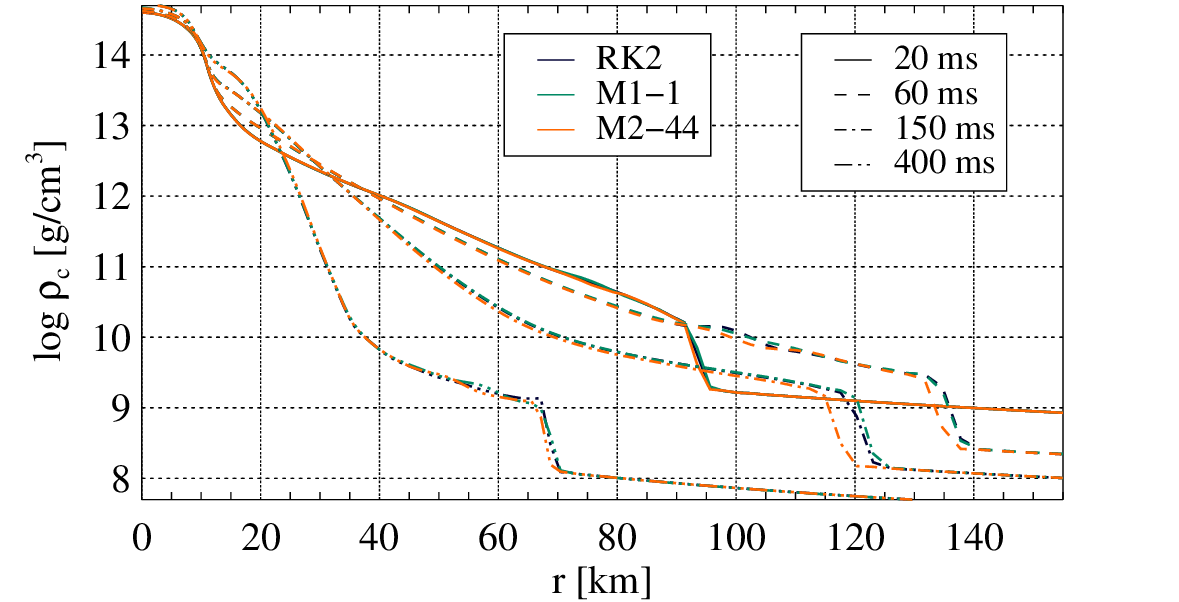}
      \includegraphics[width=0.6\linewidth]{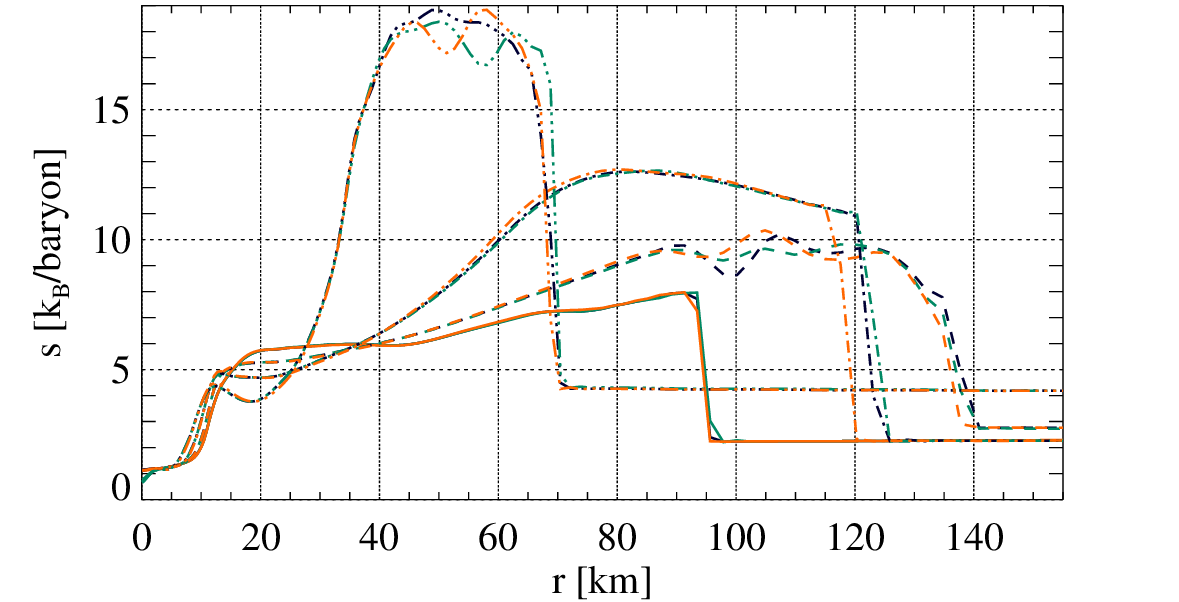}
      \includegraphics[width=0.6\linewidth]{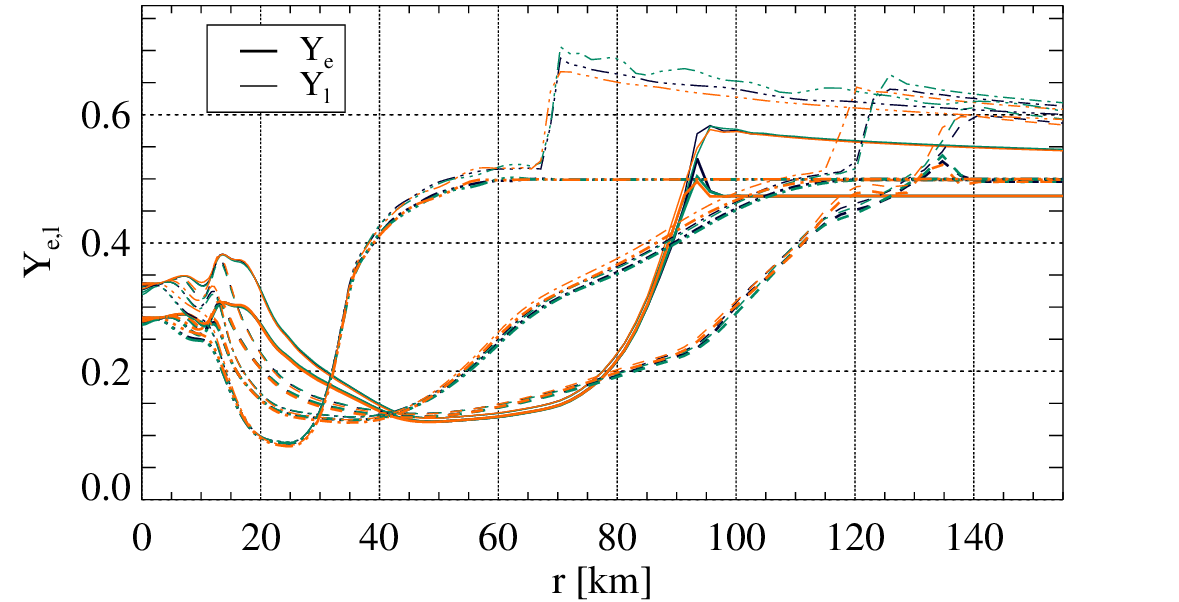}
      \includegraphics[width=0.6\linewidth]{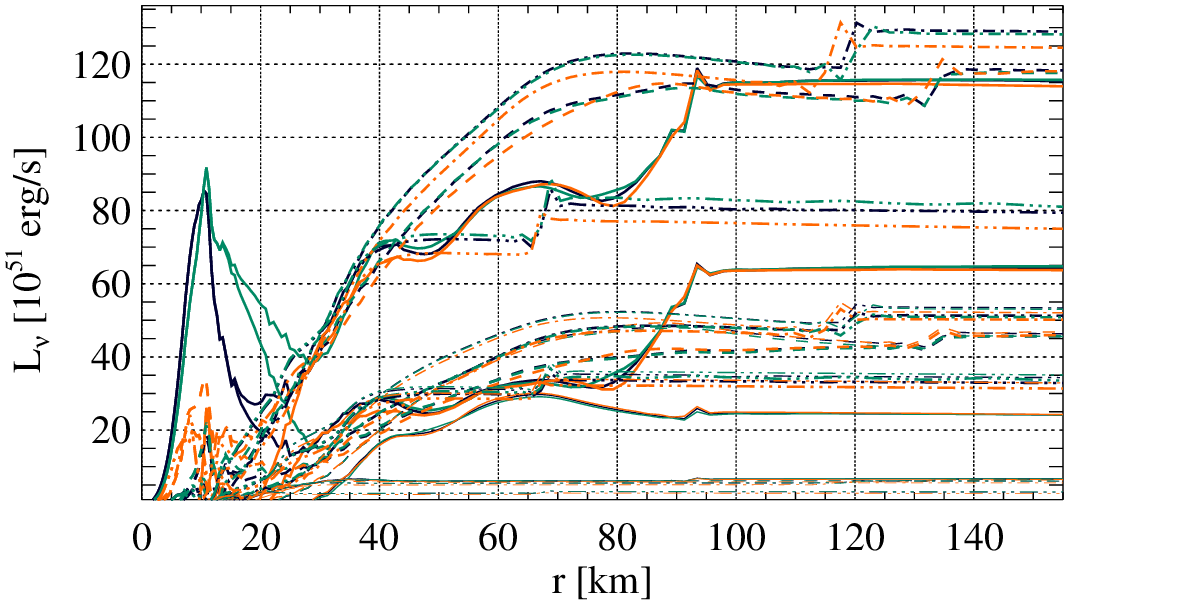}
  \caption{
    Radial profiles of selected models at a few times after bounce, as given in the legend in the top panel.
    Top panel: mass density.  
    Second panel: specific entropy.
    Third panel: electron and lepton fractions (distinguished by line thickness, see legend).
    Bottom panel: total neutrino luminosities of all flavors.
  }
  \label{Fig:profiles}
\end{figure}

\begin{table}
  \centering
  \begin{tabular}{l|cc|l}
    \hline\hline
    model & $a$ & $b$ & result
    \\
    \hline
    M1-1 & $0$ & $0$ & $\checkmark$
    \\
    M1-2 & $1/2$ & $1/2$ & $\times$
    \\
    M1-3 & $0$ & $1/2$ & $\times$
    \\
    M1-4 & $1/2$ & $0$ & $\bigtriangleup$
    \\
    \hline\hline
  \end{tabular}
  \caption{List of first order MIRK simulations performed. The first three columns give the name of the simulation, and the values of the parameters $a$ and $b$, respectively. In the last column, the symbols $\checkmark$, $\bigtriangleup$, and $\times$ indicate simulations that ran stably and with correct results, simulations that ran stably into the post-bounce phase but gave wrong results, and simulations that turned unstable when the core reached optically thick conditions, respectively.}
  \label{Tab:mirk1}
\end{table}

\subsubsection{First order MIRK numerical simulations}

The first order MIRK scheme has two free parameters, $a$ and $b$. We compare the four combinations of setting them to zero and to a
non-zero value of $1/2$ (see \tableref{Tab:mirk1}). Simulation M1-1
with $a=b=0$ satisfies the correct optically thick limit. It produces a stable simulation whose results are very close to those of the reference simulation, both in terms of the global evolution shown in \figref{Fig:rsh} and in terms of the radial profiles of \figref{Fig:profiles}. The density profiles at representative epochs during the evolution (top panel) are almost identical to those of model RK2. The PNS at the center as well as the surrounding region of decreasing density do not show any notable differences between the two simulations. The only small discrepancies appear right at the shock wave where $\rho$ falls by about an order of magnitude over a few km. The entropy and the electron/lepton fractions (second and third panels) are more sensitive than the density to the details of the neutrino treatment. Nevertheless, there are only minor deviations of model M1-1 from RK2. Apart from the shock wave, we only find small differences in the precise pattern of oscillations in the entropy behind the shock wave at late times ($t_{\mathrm{pb}}= 400 \,\mathrm{ms}$). If anything, the MIRK simulations might be able to resolve the shock wave more sharply. Furthermore, there is a minor offset of $Y_l$ outside of the shock wave. This deviation turns out to be connected to the neutrinos, not the matter, as we find a similar offset in the profiles of the neutrino luminosities (bottom panel) exterior to the shock. Among the neutrinos, we point out the relatively pronounced temporal fluctuations of the heavy lepton species, $\nu_X$, in particular of its mean energy, which we attribute to the fact that these neutrinos are generated and absorbed only via the relatively subdominant pair processes. Thus, they tend to bear the imprint of fluctuations at their production site at larger radii to a higher degree than the electron type neutrinos. In any case, the differences between the first order MIRK run and the reference solution are entirely within the margins of uncertainty of the latter alone.

\begin{figure}[htbp!]
  \centering
      \includegraphics[width=0.6\linewidth]{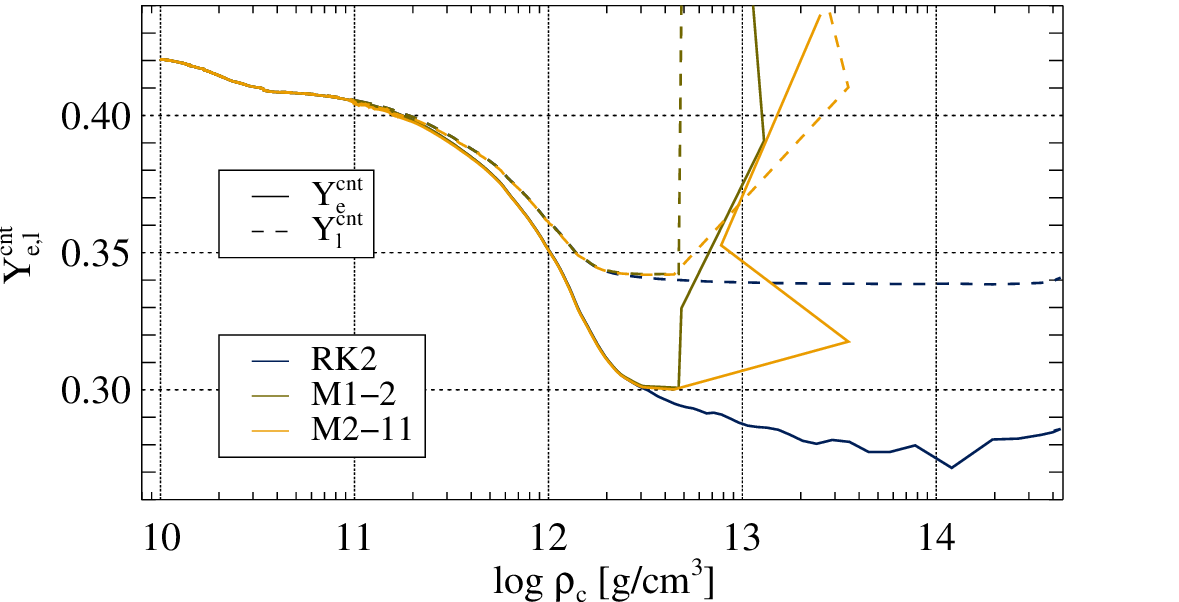}
      \includegraphics[width=0.6\linewidth]{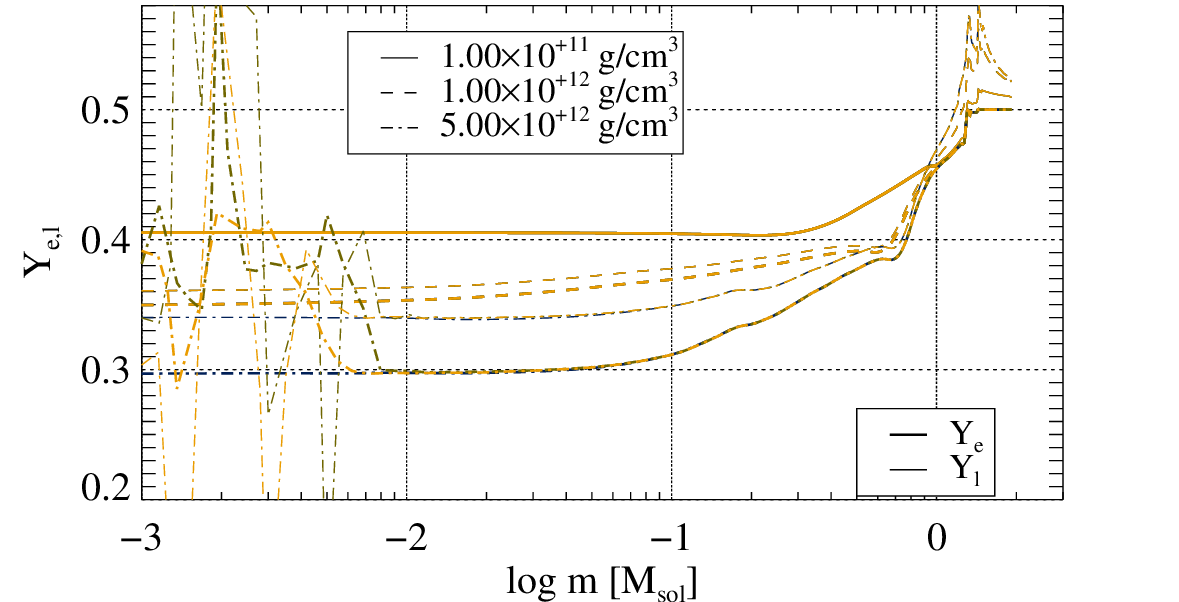}
  \caption{Comparison of unstable simulations, as indicated in the legend, to the reference simulation. 
    Top panel: evolution of the central electron and lepton fractions as function of central density during collapse.
    Bottom panel: profiles of $Y_e$ and $Y_l$ as functions of Lagrangian mass coordinates at the times at which the central density assumes the three values indicated in the legend.}
  \label{Fig:fails}
\end{figure}

The correct limit in the optically thick limit of the momentum equation is also a crucial requisite for the stability of the simulations. Models M1-2 and M1-3 with $(a,b)=(1/2,1/2)$ and $(0,1/2)$, respectively, which do not satisfy the asymptotically correct behavior (\ref{eq:opthlim1-f}) for all, smooth and non-smooth, initial data, turn unstable once the core becomes optically thick at a central density $\rho_{\mathrm{c}} \gtrsim 3 \times 10^{12} \, \gccm$ (see \figref{Fig:fails}). The instability appears first in the form of strong fluctuations near the origin that spread outward and lead to a termination of the simulation before the bounce can occur.

\begin{figure}[htbp!]
  \centering
      \includegraphics[width=0.6\linewidth]{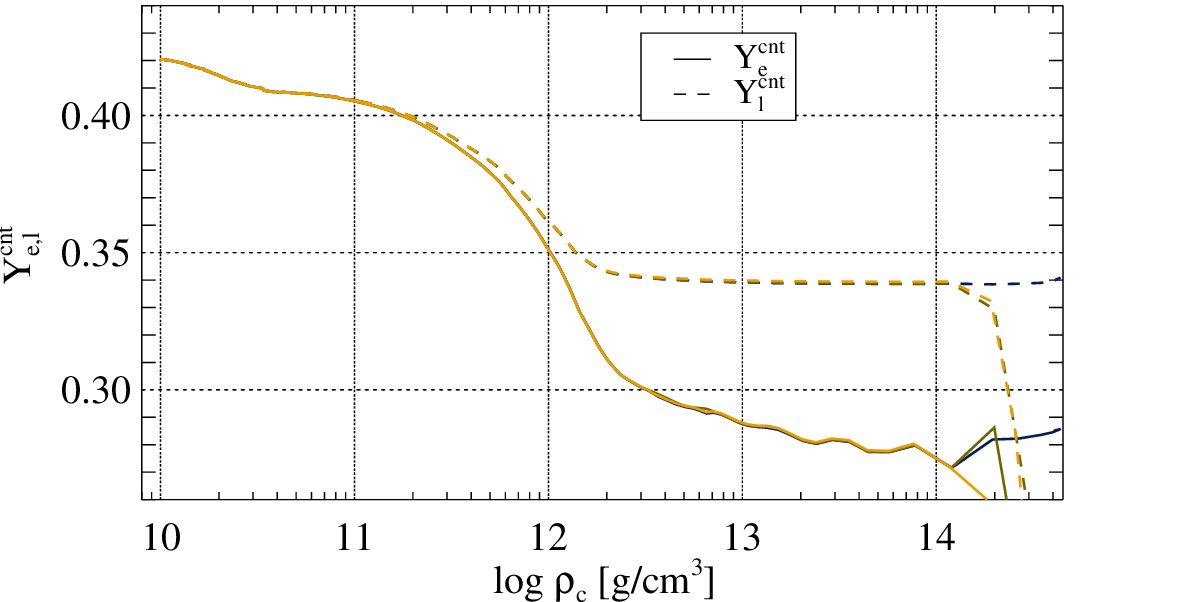}
      \includegraphics[width=0.6\linewidth]{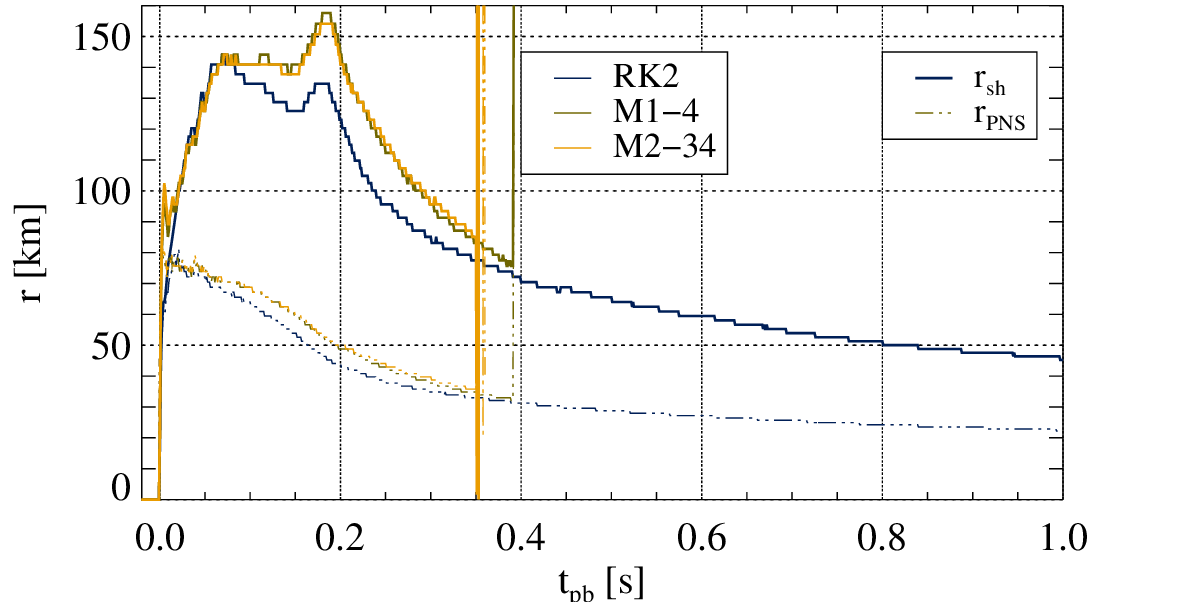}
      \includegraphics[width=0.6\linewidth]{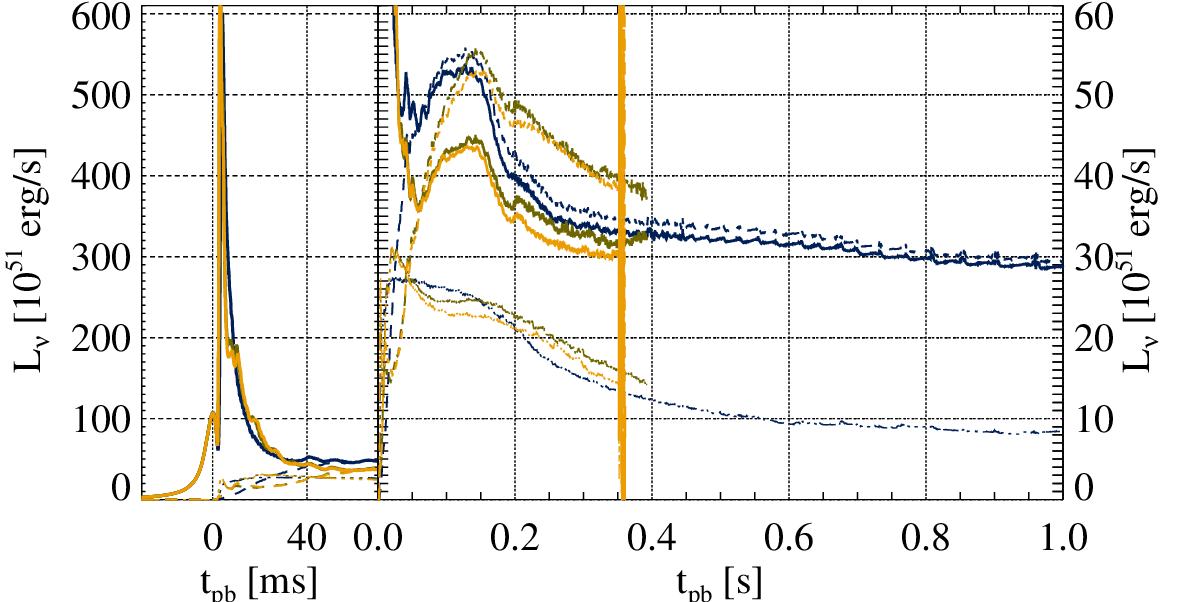}
      \includegraphics[width=0.6\linewidth]{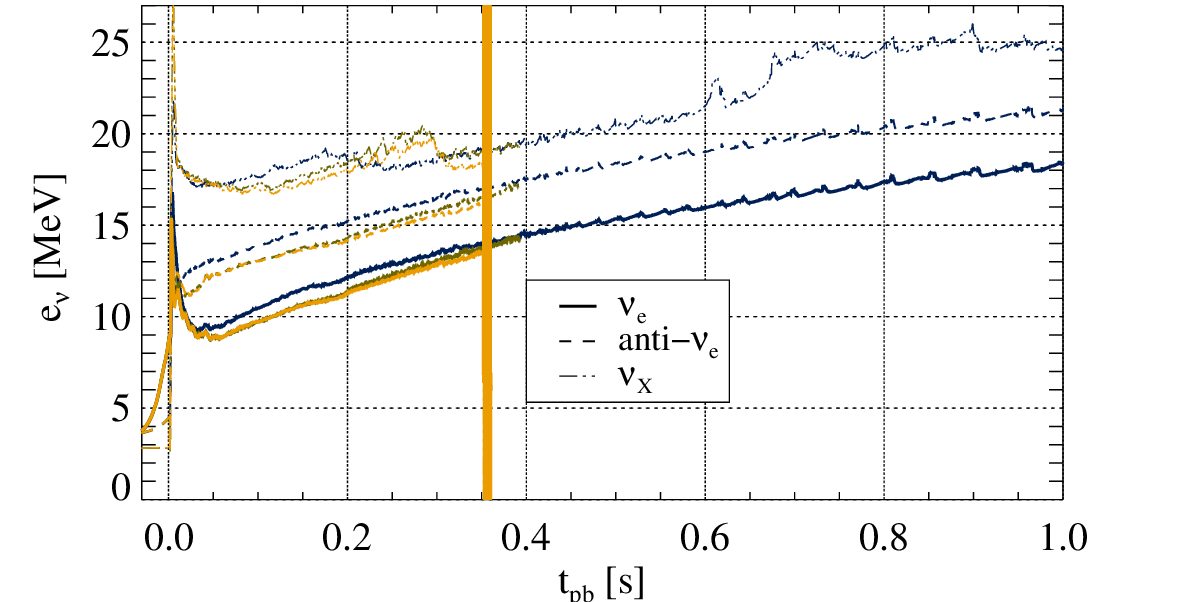}
  \caption{Same as \figref{Fig:rsh}, but comparison of simulations that evolve stably beyond bounce but produce incorrect results, as indicated in the legend, to the reference simulation.}
  \label{Fig:semifails}
\end{figure}

\begin{figure}[htbp!]
      \includegraphics[width=0.6\linewidth]{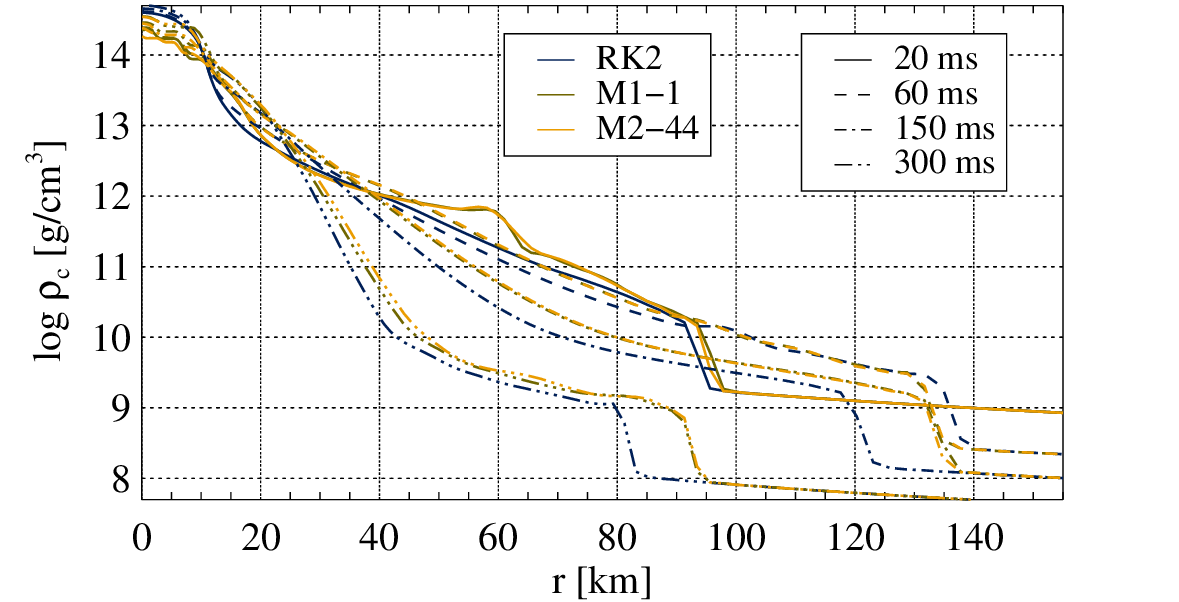}
      \includegraphics[width=0.6\linewidth]{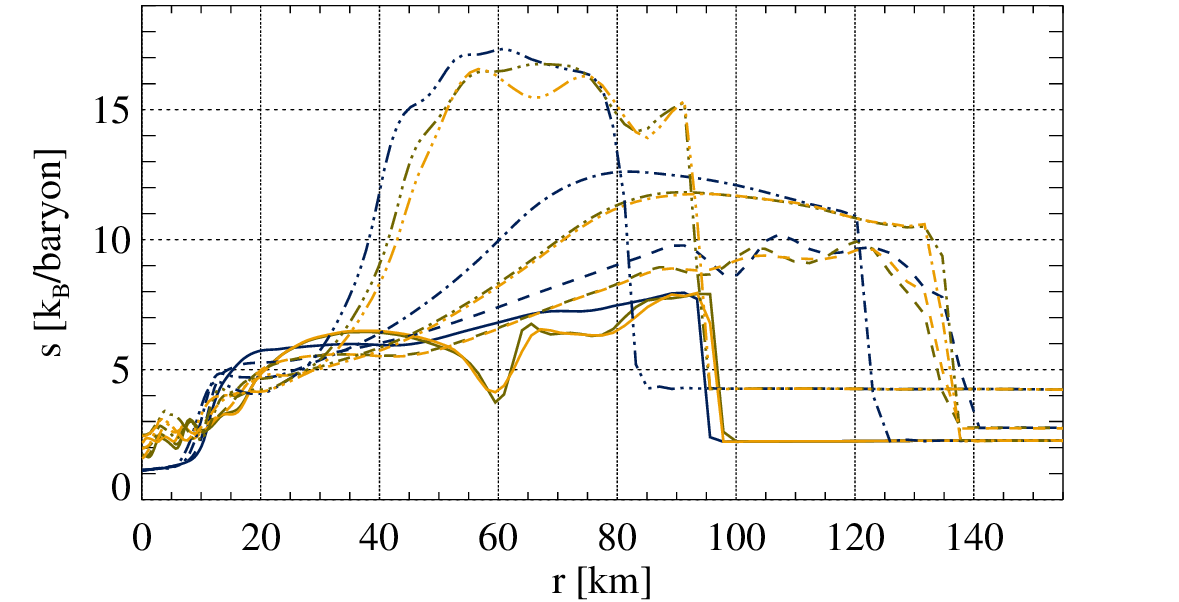}
      \includegraphics[width=0.6\linewidth]{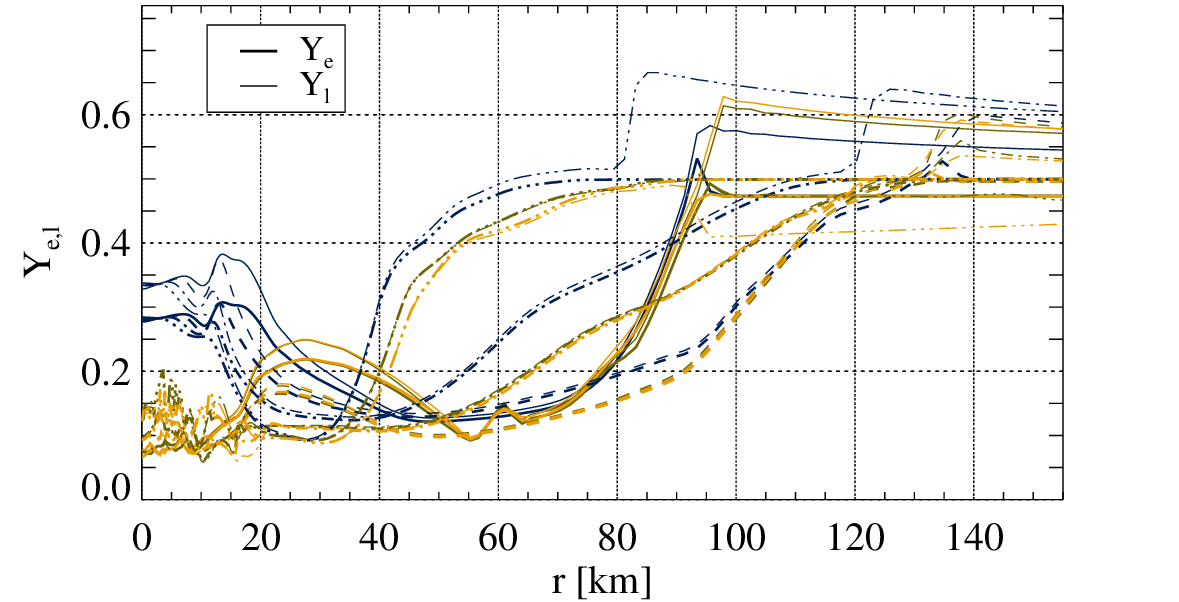}
      \includegraphics[width=0.6\linewidth]{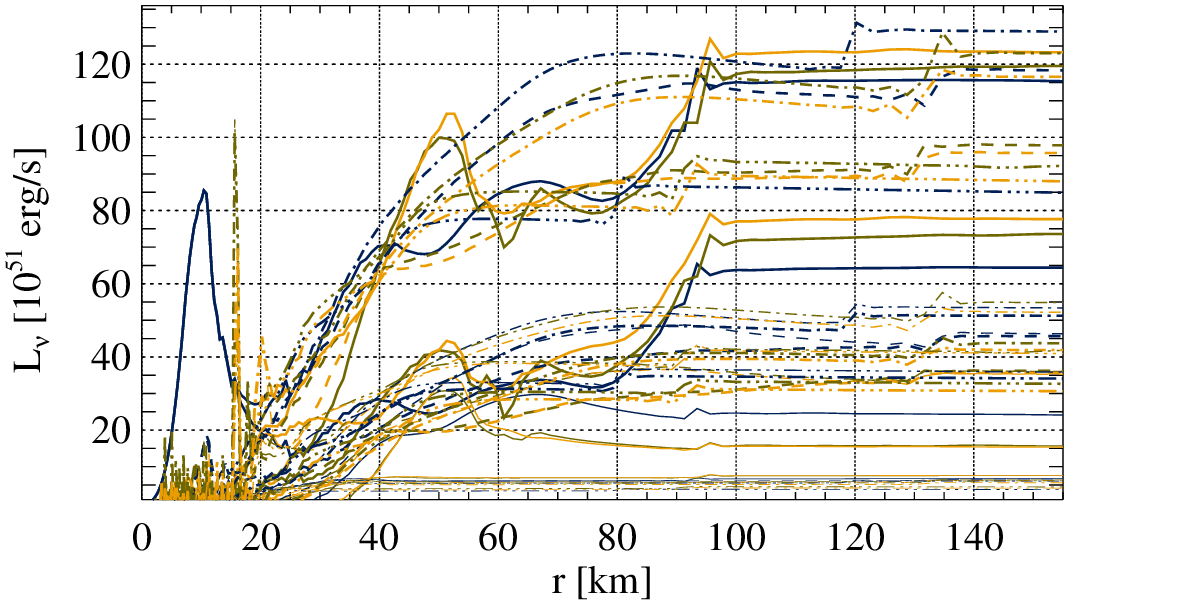}
  \centering
  \caption{Same as \figref{Fig:profiles}, but comparison of simulations that evolve stably beyond bounce but produce incorrect results, as indicated in the legend, to the reference simulation.}
  \label{Fig:semifails-prof}
\end{figure}

Model M1-4 with $(a,b) = (1/2,0)$ evolves stably and correctly through collapse and until immediately before bounce (see \figref{Fig:semifails}). The evolution of the central electron and lepton fractions agrees well with the reference model. After bounce, however, differences between the two models appear. Most notably, the central values of $Y_{e,l}$ do not stabilize at the levels they reached during neutrino trapping, but decrease further (note the drop of the two green lines in the top panel of \figref{Fig:semifails} for $\rho_{\mathrm{c}} \gtrsim 2 \times 10^{14} \, \gccm$). After about 30 ms more, they reach a minimum around $Y_{e,l} \approx 0.065$, i.e., far below the correct values. Unlike in the reference case, the two variables are almost equal at $r = 0$ throughout the entire post-bounce evolution (see third panel of \figref{Fig:semifails-prof}), i.e., the net neutrino lepton number is close to zero.  Additionally, the center of the PNS is hotter at almost twice the entropy of that of RK2 (second panel).

Discrepancies between the models are present in the $\nu_e$ burst, mostly in the form of larger fluctuations in all three flavors. Afterwards model M1-4 emits a considerably lower $L_{\nu_e}$ and higher $L_{\bar{\nu}_e}$ than RK2. The mean neutrino energies lie below the ones of RK2.

Additional differences appear in the structure of the core. Until $t_{\mathrm{pb}} \approx 20 \, \mathrm{ms}$, the shock wave transiently expands faster than in RK2. This phase is characterized by the appearance of a bump in density (see top panel of \figref{Fig:semifails-prof}, $t_{\mathrm{pb}}$, at $r \approx 60 \, \mathrm{km}$) absent from the reference case and marked differences in the entropy and $Y_{e,l}$ profiles (second and third panels). Whereas the shock wave starts to recede in the reference model after it reached its furthest expansion at $t_{\mathrm{pb}} \approx 70 \, \mathrm{ms}$, it stays in M1-4 at the same radius for another 90 ms before expanding up to $r_{\mathrm{sh}} \approx 158 \, \mathrm{km}$ and retreating only thereafter. The main differences between the models are found in the PNS, which is less dense with a shallower $\rho$ profile, hotter and neutron-richer in M1-4 than in RK2. In the surrounding hot bubble, the differences are less pronounced, but still far larger than the ones between the reference model and M1-1. Finally, a numerical instability develops in the PNS after almost 400 ms of post-bounce evolution.

To summarize, our MIRK scheme is able to reproduce the results of the reference simulations stably and correctly if the parameters $a$ and $b$ for the energy and momentum equations, respectively, are chosen such that they satisfy the correct limit in the optically thick regime. Using a different parameter in the momentum equation, i.e.,
$b\neq 0$ makes the simulations unstable once neutrinos are trapped by scattering reactions. The choice $b=0$, but $a \neq 0$, i.e., obeying the constraints in the momentum, but not the energy equation, cures this instability, but results in incorrect results once, near bounce, the emission/absorption reactions become stiff as well. The simulation can continue for several 100 ms thereafter, but the PNS properties are wrong and eventually a numerical instability ensues.

\subsubsection{Second order MIRK numerical simulations}

\begin{table}[htbp!]
  \centering
  \begin{tabular}{l|cccc|l}
    \hline\hline
    model & $a$ & $a'$ & $b$ & $b'$ & result
    \\
    \hline
    M2-11 & $+1/2$ & $-1/4$ & $+1/2$ & $-1/4$ &  $\times$
    \\
    M2-12 & $+1/2$ & $-1/4$ & $+1/2$ & $+1/4$ &  $\times$
    \\
    M2-13 & $+1/2$ & $-1/4$ & $-1/2$ & $-3/4$ &  $\times$
    \\
    M2-14 & $+1/2$ & $-1/4$ & $-1/2$ & $-9/4$ &  $\bigtriangleup$
    \\
    \hline
    M2-21 & $+1/2$ & $+1/4$ & $+1/2$ & $-1/4$ &  $\times$
    \\
    M2-22 & $+1/2$ & $+1/4$ & $+1/2$ & $+1/4$ &  $\times$
    \\
    M2-23 & $+1/2$ & $+1/4$ & $-1/2$ & $-3/4$ &  $\times$
    \\
    M2-24 & $+1/2$ & $+1/4$ & $-1/2$ & $-9/4$ &  $\bigtriangleup$
    \\
    \hline
    M2-31 & $-1/2$ & $-3/4$ & $+1/2$ & $-1/4$ &  $\times$
    \\
    M2-32 & $-1/2$ & $-3/4$ & $+1/2$ & $+1/4$ &  $\times$
    \\
    M2-33 & $-1/2$ & $-3/4$ & $-1/2$ & $-3/4$ &  $\times$
    \\
    M2-34 & $-1/2$ & $-3/4$ & $-1/2$ & $-9/4$ & $\bigtriangleup$
    \\
    \hline
    M2-41 & $-1/2$ & $-9/4$ & $+1/2$ & $-1/4$ &  $\times$
    \\
    M2-42 & $-1/2$ & $-9/4$ & $+1/2$ & $+1/4$ &  $\times$
    \\
    M2-43 & $-1/2$ & $-9/4$ & $-1/2$ & $-3/4$ &  $\times$
    \\
    M2-44 & $-1/2$ & $-9/4$ & $-1/2$ & $-9/4$ & $\checkmark$
    \\
    \hline
    M2-44-1 & $-1/4$ & $-25/8$ & $-1/4$ & $-25/8$ & $\checkmark$
    \\
    M2-44-2 & $-1/16$ & $-289/32$ & $-1/16$ & $-289/32$ & $\checkmark$
    \\
    M2-44-3 & $-1/2$ & $-9/4$ & $-1/4$ & $-25/8$ & $\checkmark$
    \\
    M2-44-4 & $-1/4$ & $-25/8$ & $-1/2$ & $-9/4$ & $\checkmark$
    \\
    \hline
    M2-51 & $-1/2$ & $-9/4$ & $3/4$ & $1/24$ & $\times$
    \\
    M2-52 & $-1/2$ & $-9/4$ & $3/4$ & $-1/8$ & $\times$
    \\
    M2-53 & $3/4$ & $1/24$ & $3/4$ & $1/24$ & $\times$
    \\
    M2-54 &$3/4$ & $-1/8$ &  $-1/2$ & $-9/4$ & $\bigtriangleup$
    \\
    M2-55 &$3/4$ & $1/24$ &  $-1/2$ & $-9/4$ & $\checkmark$
    \\
    \hline\hline
  \end{tabular}
  \caption{List of second order MIRK simulations performed. The first five columns give the name of the simulation, the values of the parameters $a$, $a'$, $b$, and $b'$, respectively. The symbols in the last column have the same meaning as in \tableref{Tab:mirk1}.}
  \label{Tab:mirk2}
\end{table}

We perform a series of simulations using the second order MIRK scheme and explore the evolution for various combinations of the four parameters $a, a', b, b'$ (see \tableref{Tab:mirk2}). The basic set of simulations consists of the 16 models M2-11, ..., M2-44, in which we set, following the possible choices introduced in \secref{sec:num}, $$a\in \{1/2, -1/2\}$$ and $$a' = \frac{a - 1}{2} \in \{-1/4, -3/4\}$$ or $$a' = \frac{(1-a)^2}{2a} \in \{1/4, -9/4\},$$ and analogously for $b$ and $b'$ (see \tableref{Tab:aapbbp} for the computation of these values based on the two possible choices previously mentioned). The nomenclature of the models is given by the following systematic scheme: the last two digits of the generic model name M2-AB indicate the values of the parameters $a$ and $b$.  Indices $\mathrm{A} = 1, 2, 3, 4$, stand for values
\begin{gather*}
    (a, a') = \left( a, \frac{a - 1}{2} \right) =(1/2,-1/4),\\ (a, a') = \left( a, \frac{(1-a)^2}{2a} \right) = (1/2,1/4),\\
    (a, a') = \left( a, \frac{a - 1}{2} \right) = (-1/2, -3/4),\\ (a, a') = \left( a, \frac{(1-a)^2}{2a} \right) = (-1/2,-9/4), 
\end{gather*}
respectively, and analogously for index $\mathrm{B}$ and parameters $(b,b')$.

\begin{table}
  \centering
  \begin{tabular}{c|cc}
    \hline \hline
    $x$ & $\frac{x-1}{2}$ & $\frac{(1 -x)^2}{2x}$
    \\
    \hline
    $-1/2$ & $-3/4$ & $-9/4$
    \\
    $-1/4$ & $-5/8$ & $-25/8$
    \\
    $-1/16$ & $-17/32$ & $-289/32$
    \\
    $1/2$ & $-1/4$ & $1/4$
    \\
    $3/4$ & $-1/8$ & $1/24$
    \\
    \hline \hline
  \end{tabular}
  \caption{Values of $x = a,b$ (first column) and the corresponding values of the two choices used to compute $x' = a',b'$ (second and third columns).
  }
  \label{Tab:aapbbp}
\end{table}

We find the same three evolutionary paths as in the first order case. Most combinations result in a numerical instability at the onset of neutrino trapping, as shown for the example of model M2-11 in \figref{Fig:fails}. As in the unstable first order runs, the instability develops in the optically thick core and causes catastrophic oscillations in the electron and lepton fractions which quickly lead to a termination of the simulations.

All simulations with $$(b, b') = \left( b, \frac{(1-b)^2}{2b} \right) = (-1/2, -9/4)$$ avoid this instability, irrespective of the values of $a$ and $a'$. However, within the basic set of the 16 models M2-11 -- M2-44 only the choice $a < 0$ and $a' = (1-a)^2/2a$ produces stable and correct results that, like for M1-1, agree very well with the reference simulation both in the evolution of global quantities (\figref{Fig:rsh}) and in the profiles at specific times (\figref{Fig:profiles}). The differences with model RK2 are limited to minor details such as the width of the shock wave (second panel of \figref{Fig:profiles}) or a small offset in the neutrino luminosity outside the shock wave.

Models M2-41, M2-42, and M2-43 with $(b,b') = (-1/2,-9/4)$ and $(a,a') \ne (-1/2,-9/4)$ show the same behavior as model M1-4 (see \figref{Fig:semifails} and \figref{Fig:semifails-prof}). Until close to the point at which the source terms in the energy equation become stiff, they follow the reference simulation. At that point, however, they yield an incorrect PNS with too low $Y_{e,l}^{\mathrm{cnt}}$, too shallow density profiles, and too low entropy. The luminosities and mean energies show the same deviations from RK2 as found in M1-4, and all models suffer the same numerical instabilities after a time of $t_{\mathrm{pb}} \sim 300 - 400 \, \mathrm{ms}$.

We added models M2-44-1 -- M2-44-4 similar to M2-44. Their results agree well with those of model M2-44, indicating that stability and accuracy do not depend on the specific values as long as $$(a', b') = \left(\frac{(1-a)^2}{2 a}, \frac{(1-b)^2}{2b}\right)$$ and $a,b < 0$ are satisfied.

Another group of simulations, models M2-51 -- M2-55, probe positive values of $a$ and $b$ between $1/2$ and $1$, which according to \Eqref{eq:choiceb2a} and \eqref{eq:choiceb2} could also lead to a stable evolution. However, we find that all simulations with $b=3/4$ are unstable. If we set, as in M2-44, $(b,b') = (-1/2, -9/4)$, we obtain a stable and correct simulation with $(a,a') = (3/4, 1/24)$ and a stable, but incorrect one with $(a,a') = (3/4, -1/8)$.

Hence, we find that, similarly to the first order schemes, the stability is set by the parameters for integrating the momentum equation: only $b' = (1 -b )^2/2b$ and $b < 0$ are stable. Among these, the ones for which the parameters for the energy equation fulfil the  constraint $a' = (1-a)^2/2a$ and $a<0$ or $1/2<a<1$ are also correct.

Let us point out again that, although the stability analysis for the MIRK methods is done at the stiff limit, the simulations performed in this subsection and the previous one go through the transition from the non-stiff to the stiff regime. The numerical results agree with the expected results from the theoretical stability analysis performed.

\section{Conclusions}
\label{sec:conclusions}

We have derived a Minimally Implicit Runge-Kutta method for $M_1$ equations for neutrino transport. We use it to treat the neutrino matter interaction terms describing reactions such as absorption, emission, and scattering in an operator-split manner separately from the (hyperbolic) transport terms of an $M_1$ method. In general, the stiffness of the interaction terms in the optically thick regime poses a stability problem for their time integration. The problem can be overcome by fully implicit methods, but these can be very costly because of the complex dependence of the reaction rates on the neutrino fields and the thermodynamic state of the matter. We propose a simplified approach that reduces the use of implicit terms to the minimum required for stability by evaluating the opacities and the thermodynamics explicitly. This choice makes that the resulting scheme takes a form similar to that of an explicit method. The first order method is a straightforward modification of an explicit scheme with an effective, reduced time step that guarantees stability. The method implemented in this context in the neutrino-hydrodynamics code Alcar \citep{just2015} can be viewed as a particular case of the MIRK method. Here we give a mathematical framework, providing arguments based on the behavior of the evolved variables and stability criteria at the stiff limit, and a generalization to higher order methods that retains the simplicity of the first order one. Similar schemes were already applied in \citep{cordero2023} for the resistive relativistic magnetohydrodynamic equations.

The second order method depends on two numerical parameters. We demonstrate that these parameters can be chosen in such a way that they satisfy an algebraic condition that guarantees the correct optically thick limit for the source terms for neutrino energy and momentum. 

We have tested the MIRK scheme in a serious of tests with an increasing degree of complexity. The scheme is able to recover the analytic solution of the diffusion of radiation through a uniform and constant background. It also performs well in a toy model of the neutrino emission in a PNS in which we allow for a feedback on the internal energy of the gas by thermal emission and absorption of neutrinos. The most demanding tests case to which we subjected the scheme, after implementing it in the Alcar supernova code, is a spherically symmetric simulation of core collapse with full coupling to the hydrodynamics of the gas and a consistent treatment of the microphysics (equation of state, neutrino reactions). In this case, the new time integrator gives stable and accurate results. If, on the other hand, the stability conditions we derived are violated, the simulations become unstable once the core turns optically thick. We note that this simulation differs from the class of state-of-the-art supernova simulations mainly by its dimensionality. As this factor is independent of the method used to treat the spatially local source terms of neutrino-matter interaction, we consider the setup an entirely meaningful test of the new scheme in a system with a reasonable degree of realism.

Our scheme is simple and efficient and can be used in a wide range of similar applications. Examples can be found in contexts with rarefied gases \citep{koellermeier2022rar}, shallow water equations, \citep{koellermeier2022shallow} or force-free electrodynamics in General Relativity \citep{Mahlmann2021}. There, stiff terms appear in the conservative laws coming from the corresponding scenarios. We note that the scheme, as presented here, is adapted to the case of source terms that describe the relaxation of the neutrino moments to a (thermal) equilibrium state. For the moment, the interactions that have to be included in models of CCSNe or similar systems, such as inelastic scattering or pair processes, are treated in an operator-split approach in a separate step. How to include such terms in a MIRK scheme remains the topic of future work, as will be the combination with a well-balanced method (see \citep{castro2020well}) to manage the fluxes in order to preserve stationary solutions.

\ack{This research was partially supported by the Perimeter Institute for Theoretical Physics through the Simons Emmy Noether program. Research at Perimeter Institute is supported by the Government of Canada through the Department of Innovation, Science and Economic Development and by the Province of Ontario through the Ministry of Research and Innovation.}

\funding{The authors acknowledge support by the Spanish Agencia Estatal de Investigación through the Grants No. PGC2018-095984-B-I00, PID2021-125485NB-C21, PID2021-127495NB-I00, PID2022-136828NB-C43 and PID2024-159689NB-C21 funded by MCIN/AEI/10.13039/501100011033 and by the European Union  “NextGenerationEU" as well as “ESF Investing in your future”, by the Generalitat Valenciana through the Prometeo program for excellent research groups Grant No. PROMETEO/2019/071, CIPROM/2022/49, CIPROM/2022/13, Grant No. ACIF/2019/169 - European Social Fund and the Astrophysics and High Energy Physics program Grants No. ASFAE/2022/0003 and ASFAE/2022/026 funded by MCIN and the European Union NextGenerationEU (PRTR-C17.I1), and by the European Horizon Europe staff exchange (SE) programme HORIZON-MSCA-2021-SE-01 Grant No. NewFunFiCO-101086251. MO was supported by the Ramón y Cajal programme of the Agencia Estatal de Investigación (RYC2018-024938-I).
}

\sloppy
\bibliographystyle{apsrev4-2}
\bibliography{article}

\end{document}